\documentclass[iop]{emulateapj}

\usepackage{color}
\usepackage{natbib}

\slugcomment{Submitted to ApJ}

\shorttitle{3D AMR simulations of the gas cloud G2}
\shortauthors{Schartmann et al.}

\begin{document}


\title{3D adaptive mesh refinement simulations of the gas cloud G2 born within the disks of
  young stars in the Galactic Center}


\author{M.~Schartmann\altaffilmark{1}\altaffilmark{2}\altaffilmark{3}, A.~Ballone\altaffilmark{1}\altaffilmark{2}, A.~Burkert\altaffilmark{1}\altaffilmark{2}\altaffilmark{4},
S.~Gillessen\altaffilmark{2}, R.~Genzel\altaffilmark{2},
O.~Pfuhl\altaffilmark{2}, F.~Eisenhauer\altaffilmark{2},
P.~M.~Plewa\altaffilmark{2}, T.~Ott\altaffilmark{2},
E.~M.~George\altaffilmark{2}, M.~Habibi\altaffilmark{2}}   

\email{mschartmann@swin.edu.au}

\altaffiltext{1}{Universit\"ats-Sternwarte M\"unchen, Scheinerstra\ss e 1, D-81679 M\"unchen, Germany}
\altaffiltext{2}{Max-Planck-Institut f\"ur extraterrestrische Physik, Postfach 1312, Giessenbachstr., D-85741 Garching, Germany}
\altaffiltext{3}{Centre for Astrophysics and Supercomputing, Swinburne
  University of Technology, Hawthorn, Victoria 3122, Australia}
\altaffiltext{4}{Max-Planck-Fellow}

\begin{abstract}
The dusty, ionized gas cloud 
  {\it G2} is currently
passing the massive black hole in the Galactic Center at a
distance of roughly 2400 Schwarzschild radii.
We explore the possibility of a starting point of the cloud
within the disks of young stars.  
We make use of the large amount of new observations 
in order to put constraints on G2's origin. Interpreting the observations as a diffuse cloud of
gas, we employ three-dimensional hydrodynamical adaptive mesh
refinement (AMR) simulations with
the {\sc PLUTO} code and do
a detailed comparison with observational data.  
The simulations presented in this work update our previously obtained results in multiple ways:
(1) high resolution three-dimensional hydrodynamical AMR simulations
are used, (2) the cloud follows the updated orbit based on the
Brackett-$\gamma$ data, 
(3) a detailed comparison to the observed high-quality position-velocity diagrams
and the evolution of the total Brackett-$\gamma$ luminosity is done.
We concentrate on two unsolved problems of the diffuse cloud
scenario: the unphysical formation epoch only shortly before the first
detection and the too steep Brackett-$\gamma$ light curve obtained in simulations, whereas the
observations indicate a constant Brackett-$\gamma$ luminosity between
2004 and 2013.  
For a given atmosphere and cloud mass, we find a
consistent model that can explain both, the observed Brackett-$\gamma$ light curve and
the position-velocity diagrams of all epochs. Assuming initial
pressure equilibrium with the atmosphere, this can be reached for a starting date
earlier than roughly 1900, which is close to apo-center and well within the disks of young stars. 

\end{abstract}




\section{Introduction}
\label{sec:introduction}

The density distribution and
kinematics of objects in the Galactic Center are strongly affected
by tidal forces due to the central massive black hole (BH), feedback processes from the cluster of
high-mass stars in the direct vicinity of Sgr~A*, and the central gas accretion flow.
A very prominent example is the recently discovered dusty, ionized gas
cloud, G2. It is on an extremely elliptical orbit ($e=0.98$), bringing it as close as 2400 Schwarzschild radii 
to the central massive BH. 
The fortuitous detection only a few years before G2's pericenter
passage in early 2014 (during its roughly 400\,yr orbital period),
enabled us to get a live view of the unfolding tidal disruption.
This is best seen in position-velocity diagrams constructed from
the Brackett-$\gamma$ emission (but also in other gas tracer lines)
detected with the SINFONI \citep{Eisenhauer_03,Bonnet_04} instrument at the VLT 
\citep[][orbital properties taken from
\citealp{Gillessen_13b}]{Gillessen_12, Gillessen_13a,
  Gillessen_13b, Pfuhl_15}
or OSIRIS \citep{Larkin_06} at the Keck telescopes \citep{Phifer_13}. 
The expected rising signature of
tidal disruption in total Brackett-$\gamma$ emission has only been seen very close to the nominal time of 
pericenter passage, following a long plateau 
($L_{\mathrm{Br}\gamma}\approx2\times10^{-3}\,L_{\odot}$) and hence poses a challenge
to available models. 

In contrast to the clear disruption signature of the Brackett-$\gamma$
emission, recent L' band observations (targeting the dust content) using the
near-infrared camera NIRC2 and the Keck II laser guide star adaptive
optics system \citep[LGSAO,][]{Wizinowich_06,vanDam_06} show L' band diameters of G2 smaller than 260
astronomical units, with a constant magnitude of 14 between 2005 and
2014, even after pericenter passage \citep{Witzel_14}.
These observations give important constraints for
theoretical models and might help to differentiate between 
possible models. 
Dust, however, makes up only a tiny fraction
of the total mass of G2. \citet{Pfuhl_15} find a dust mass of roughly
$10^{-12}\,M_\mathrm{\odot}$.
This is consistent
with the dust content of a second gas cloud (G1) in the Galactic
Centre \citep{Ghez_05}, which evolves on a similar orbit but precedes
G2 by roughly 13 years.
Due to this very small dust-to-gas ratio, we 
do not expect the dust component to affect the dynamics and
distribution of the gas and neglect its contribution in this article. 

Further very recent near-infrared observations of G2 have been reported by
\citet{Valencia_15}. They find no blue-shifted emission in their 
February to May 2014 SINFONI data set, but only red-shifted emission and
vice versa for the data after May 2014. 
No significant line broadening of the detected 
red-shifted  Brackett-$\gamma$ line with respect to their 2013 data
set was detected. This is interpreted as G2 having passed pericenter
in 2014.39 and as an indication for the compactness of the source.

With a SINFONI data set that reaches the highest signal-to-noise ratio of all
available data, \citet{Pfuhl_15} come to a different conclusion: with
time, the redshifted emission fades and the blue-shifted
Brackett-$\gamma$ emission gradually appears and starts to dominate in mid-2014.
Evidence is found
for a connection of the G2 cloud and the G1 cloud, indicating that a
drag force might lead to deviations from a purely Keplerian orbit.
This might allow future observations to constrain properties of the hot accretion flow
\citep{Pfuhl_15,McCourt_15b}. 
Having reached its pericenter passage roughly 13 years
earlier, G1 might give us a preview of what will happen to G2 in the
coming years. G1 and the presence of a tail following G2 suggest 
that they are part of a more extended stream of
gas pointing towards Sgr~A*.

Many monitoring campaigns of Sgr~A* at various wavelengths 
have been undertaken since the detection of
G2 in 2012, but none of them has found 
unambiguous evidence for a change in the
flux density or the activity of Sgr A*
which could be attributed to the interaction of G2 gas with the
central black hole \citep[][and references therein]{Bower_15,Park_15}.
After carefully analysing {\it XMM Newton} and
{\it Chandra} observations, \citet{Ponti_15} found that the bright or very bright
X-ray flare luminosity of Sgr~A* increased by a factor of 2 to 3 between
2013 and 2014. A factor of approximately 9 increase of the bright or very
bright flaring rate is reported, which started roughly 6 months after the nominal
pericenter passage of G2.  
Given the not very well known power spectrum of the X-ray
flare emission, this could also be caused by the increased monitoring frequency
triggered by the G2 detection \citep{Ponti_15}.

As an ideal testbed for the investigation of physical processes in galactic nuclei, this
event led to immediate theoretical interest. 
\citet{Burkert_12} investigated the most
important physics involved and 
possible formation mechanisms for the G2 cloud.
Models can be largely separated into two main categories: (1) diffuse gas clouds
and (2) gas clouds containing compact sources. 
The first class could be interpreted as debris from stellar wind interactions
\citep{Gillessen_12, Burkert_12, Cuadra_06} or condensations within a
stream of gas \citep{Guillochon_14,Pfuhl_15}. The latter scenario was
for example modelled as a photoevaporating protoplanetary / circumstellar disk
\citep{Murray_Clay_12,Miralda_Escude_12},  a mass-losing star
\citep{Meyer_12,Scoville_13,Ballone_13,Colle_14} 
or the product of a stellar
binary merger due to so-called Kozai-Lidov oscillations induced by the central massive
black hole \citep{Phifer_13,Prodan_15,Witzel_14}.

We will concentrate on the case of a diffuse gas cloud.
First two-dimensional hydrodynamical simulations of such a scenario have been
presented by \citet{Schartmann_12}. 
Assuming the gas clouds start in pressure
equilibrium, have a given gas mass and employing simple test
particle simulations, the data necessitated a comparably recent starting point of a Compact Cloud
close to the year 1995. 
But a starting point closer to apocenter is more favorable for several reasons: (1) The cloud
spends most of its lifetime in this part of the orbit. (2) It is well within the range of 
the disk(s) of young stars. 
The latter are made up of roughly 100 massive O- and Wolf-Rayet stars
distributed in a warped clockwise rotating disk 
ranging from roughly 0.05 to around
0.5\,pc \citep{Paumard_06,Bartko_09,Lu_09,Yelda_14}. 
A fraction of the stars is counter-clockwise rotating and
might be associated with a second, inclined disk \citep[e.~g.~][]{Paumard_06}. 
The cloud could then be made up of
shocked debris from the interaction of slow winds within these disks
of young stars \citep{Burkert_12}.
(3) No source of gas has been
identified in the vicinity of a starting point around the year 1995. 
Using the same initial conditions as found in \citet{Schartmann_12}, 
\citet{Anninos_12} ran 3D moving mesh hydrodynamical simulations
of the G2 cloud. Similar values were found for 
the mass transfer rate towards the center. However, no direct comparison
to observations was done.  
The alternative formation scenario which interprets the cloud as the
result of mass loss from a compact central source of gas
is presented in a companion paper (A.~Ballone et al.~, 2015, in
preparation), together with a comparison of the two basic scenarios.

Despite the large amount of observations and theoretical
investigations, the nature of the G2 cloud and
its trailing component G2t is still a mystery and many questions
remain unanswered, e.~g.~
\begin{enumerate}
\item Is it a pure gas cloud or does it hide a mass-losing source? 
\item How did it end up on such a high eccentricity orbit? 
\item What is the physical origin of the plateau in the
Brackett-$\gamma$ light curve?
\item What is the physical connection between G2, its trailing
component G2t, and G1?
\end{enumerate}

The goal of this paper is to build on the knowledge gained so far
and set up three-dimensional hydrodynamical simulations to
assess the possible origin and fate of the G2 cloud. 
The most important observational constraints, as well as the updated
orbital information, will be taken into 
account to better constrain a possible initial condition in the 
framework of the compact cloud scenario and to shed light on the 
detailed evolution. Most importantly, the 3D AMR 
simulations allow us to make a much more detailed direct comparison to
available data. Such a quantitative analysis was not possible with our preliminary 2D
results, wherein the internal cloud structure could not be assessed accurately.

In Sect.~\ref{sec:num_setup} we summarize the numerical setup and the
parameter settings. The evolution of the gas density of the
clouds is presented in Sect.~\ref{sec:densevol}.
The simulations are compared to observational data in Sect.~\ref{sec:obscomp}. 
A critical discussion is presented in Sect.~\ref{sec:discussion} and the conclusions
in Sect.~\ref{sec:conclusions}.

\section{Simulation setup}
\label{sec:num_setup}

In order to derive the simplest realization of the Compact Cloud
scenario, we follow closely the basic setup already used for the simulations
described in \citet{Schartmann_12} and \citet{Ballone_13}, which we
briefly summarize here.
The initially spherical cloud evolves in the potential of the massive central BH
($M_{\mathrm{BH}}=4.31\times 10^6\,M_{\odot}$) and is embedded into a
hot atmosphere.
We model the atmosphere following the
ADAF realization presented in \citet{Yuan_03}:

\begin{eqnarray}
\label{equ:numdens_at}
n_{\mathrm{at}} = 930\,\mathrm{cm}^{-3}\,f_{\mathrm{hot}}\,\left(\frac{1.4\times 10^4 R_{\mathrm{S}}}{r}\right)^{\alpha} \\
\label{equ:temp_at}
T_{\mathrm{at}} = 1.2\times 10^8\,\mathrm{K}\,\left(\frac{1.4\times 10^4 R_{\mathrm{S}}}{r}\right)^{\beta}
\end{eqnarray}

where $n_{\mathrm{at}}$ is the number density distribution and
$T_{\mathrm{at}}$ the temperature distribution, both only depending on
the distance to the BH $r$.
$R_{\mathrm{S}}$ refers to its Schwarzschild radius, the exponents
$\alpha$ and $\beta$ are both set to one and $f_{\mathrm{hot}}\approx 1$ is a factor taking the
uncertainty of the model into account, which we set to one here.
A mean molecular weight of $\mu=0.6139$ has been assumed, typical for a gas with solar metallicity.
Following \citet{Schartmann_12}, we artificially stabilize the
atmosphere, which is unstable to convection. This is done by
additionally evolving a tracer field ($0 \leq tr \leq 1$), 
which obeys a simple advection equation and passively follows the
fluid:

\begin{eqnarray}
\frac{\partial(\rho\,tr)}{\partial t} + \mathbf{\nabla} \cdot (\rho\,tr\,\mathbf{v}) = 0,
\end{eqnarray}

where $\rho$ is the gas density, $t$ the time and $\mathbf{v}$ the
fluid velocity. 
We initially assign the cloud a value of one for this passive tracer field and the atmosphere zero. This 
allows us to distinguish 
between those parts of the atmosphere which have interacted with the cloud ($tr \geq 10^{-4}$) from those which changed
due to the atmosphere's inherent instability ($tr < 10^{-4}$). 
In those cells that fulfil the latter criterion ($tr <
  10^{-4}$)  we
  reset the density, pressure and velocity to  
the values expected in hydrostatic equilibrium.
For a more detailed description,
and a discussion of the consequences for the evolution
  of the simulations,
we refer to \citet{Schartmann_12}.
Deviating from the work presented there, we update
the simulations in two ways: 
(1) the latest observationally determined best-fit orbital solution
for the G2 cloud
is used, which is the Brackett-$\gamma$ based orbit as determined by
\citet[][see their Table~1]{Gillessen_13b} and
(2) 3D AMR calculations are employed.
Three-dimensional simulations have the main advantage that they allow
for a detailed comparison with the observed position-velocity diagrams
as well as the time evolution of the Brackett-$\gamma$ emission. 
With our previous 2D simulations, no quantitative comparison was
possible. The new 3D simulations allow for the best use of the available data in 
constraining the initial conditions for the cloud. Providing such a
best-fit hydrodynamical model for G2 in the framework of the Compact
Cloud Scenario is the main objective of this publication.

A Cartesian coordinate system spanning a volume of $-2.7\times 10^{17}$\,cm to $3.7\times 10^{16}$\,cm in $x$-direction,
$-7.8\times 10^{16}$\,cm to $7.8\times 10^{16}$\,cm in $y$-direction and
$-3.9\times 10^{16}$\,cm to $3.9\times 10^{16}$\,cm in
$z$-direction is used. The orbital plane is given by
the x-y plane.
Using up to six levels of refinement,
a spatial resolution of $3.8\times 10^{13}$cm is reached.
The spherical cloud is initially in pressure equilibrium on a clockwise orbit within the $x$-$y$-plane with the major axis along the $x$-axis and the pericenter 
of the orbit on the positive $x$-axis. The BH is located at the origin of 
our coordinate system.
Various starting positions along the observed orbit \citep{Gillessen_13b} are
tested and the implications of a resolution study are discussed, see
Table~\ref{tab:simparam} for an overview of all simulations. 
The mass of the cloud is fixed to the value estimated from
observations ($M_{\mathrm{cloud}}=1.7\times10^{28}\,$g) in \citet{Gillessen_12}. 

The simulations make use of the PLUTO code
\citep[v4.0,][]{Mignone_07,Mignone_12} to integrate the hydrodynamical
equations, for which the two-shock Riemann solver is chosen together
with a parabolic interpolation and the second-order Runge-Kutta time
integration scheme. All boundaries are set to the
hydrostatic equilibrium values expected for the hot atmosphere. 
This includes a sphere with a radius of $10^{15}$\,cm
surrounding the central massive BH.
Gas is allowed to flow from the computational domain into this central
cavity, but not vice versa. 
Only the equations of
hydrodynamics are taken into account. We exclude thermal conduction,
magnetic fields or any kind of feedback from the central source.
The AMR technique is used to enable
efficient calculations of the cloud evolution in 3D, where we refine
according to the second derivative of the density field. Apart from this,
the same numerical schemes are applied as described in
\citet{Schartmann_12}.


\begin{table*}
\begin{center}
\caption{Parameters of the hydrodynamical simulations.\label{tab:simparam}}
\begin{tabular}{lcrrrrrrrr}
\tableline\tableline
name & $\tau_{0}$\tablenotemark{a} &  $\rho_{\mathrm{cloud}}$\tablenotemark{b} & $R_{\mathrm{cloud}}$\tablenotemark{c} &
$x_{\mathrm{ini}}$\tablenotemark{d} & $y_{\mathrm{ini}}$\tablenotemark{e} & $v^x_{\mathrm{ini}}$\tablenotemark{f} &
$v^y_{\mathrm{ini}}$\tablenotemark{g} & $\Delta x$\tablenotemark{h}\\
 & {\footnotesize yr AD} & {\footnotesize $10^{-19}\,\mathrm{g}\,\mathrm{cm}^{-3}$} & {\footnotesize $10^{15}\,\mathrm{cm}$} & 
{\footnotesize $10^{16}\,\mathrm{cm}$} & {\footnotesize $10^{16}\,\mathrm{cm}$} 
& {\footnotesize $\mathrm{km}\,\mathrm{s}^{-1}$} & {\footnotesize $\mathrm{km}\,\mathrm{s}^{-1}$}
& {\footnotesize $10^{13}\,\mathrm{cm}$}\\
\tableline
1818L & $1817.9$ & 0.53 & 4.24 & -25.80 & 0.00 & 0.00 & 72.64 & 15.17\\
1850L & $1850.0$ & 0.55 & 4.19 & -25.36 & 0.73 & 88.02 & 71.37 & 15.17\\
1850M & $1850.0$ & 0.55 & 4.19 & -25.36 & 0.73 & 88.02 & 71.37 & 7.59 \\
1850H & $1850.0$ & 0.55 & 4.19 & -25.36 & 0.73 & 88.02 & 71.37 & 3.79 \\
1870L & $1870.0$ & 0.58 & 4.11 & -24.62 & 1.18 & 145.58 & 69.17 & 15.17\\
1880L & $1880.0$ & 0.61 & 4.06 & -24.11 & 1.39 & 175.84 & 67.57 & 15.17\\
1880M\tablenotemark{i} & $1880.0$ & 0.61 & 4.06 & -24.11 & 1.39 & 175.84 & 67.57 & 7.59\\
1880H & $1880.0$ & 0.61 & 4.06 & -24.11 & 1.39 & 175.84 & 67.57 & 3.79\\
1890L & $1890.0$ & 0.64 & 3.99 & -23.51 & 1.60 & 207.47 & 65.58 & 15.17\\
1900L & $1900.0$ & 0.68 & 3.91 & -22.80 & 1.80 & 240.85 & 63.12 & 15.17\\
1920L & $1920.0$ & 0.79 & 3.72 & -21.05 & 2.18 & 314.84 & 56.36 & 15.17\\
1971L & $1971.0$ & 1.75 & 2.85 & -13.96 & 2.82 & 604.22 & 12.24 & 15.17\\
1995L & $1995.0$ & 4.62 & 2.06 & -8.34 & 2.67 & 928.82 & -72.12 & 15.17\\
\tableline
\end{tabular}
\tablenotetext{1}{Start time of the simulation.}
\tablenotetext{2}{Initial density of the cloud.}
\tablenotetext{3}{Initial radius of the cloud.}
\tablenotetext{4}{Initial $x$-position of the cloud.}
\tablenotetext{5}{Initial $y$-position of the cloud.}
\tablenotetext{6}{Initial $x$-velocity of the cloud.}
\tablenotetext{7}{Initial $y$-velocity of the cloud.}
\tablenotetext{8}{Minimum cell size in $x$-, $y$- and $z$-direction.}
\tablenotetext{9}{Model 1880M is referred to as the {\it standard model}
throughout the paper.}
\tablecomments{
Note that we always start on the same orbit, however at different
locations, corresponding to different starting times.
The simulation name is made up of the starting date
  followed by a letter indicating the maximum resolution reached in
  the simulation (see last column). 1817.9 corresponds to the apo
  center of the nominal orbit and 1971 is the time when G2's orbit crosses the inner
  radius of the disk of young stars at approximately
  0.05\,pc.}
\end{center}
\end{table*}

\section{Evolution of the density distribution}
\label{sec:densevol}

\begin{figure*}
\epsscale{1.0}
\plotone{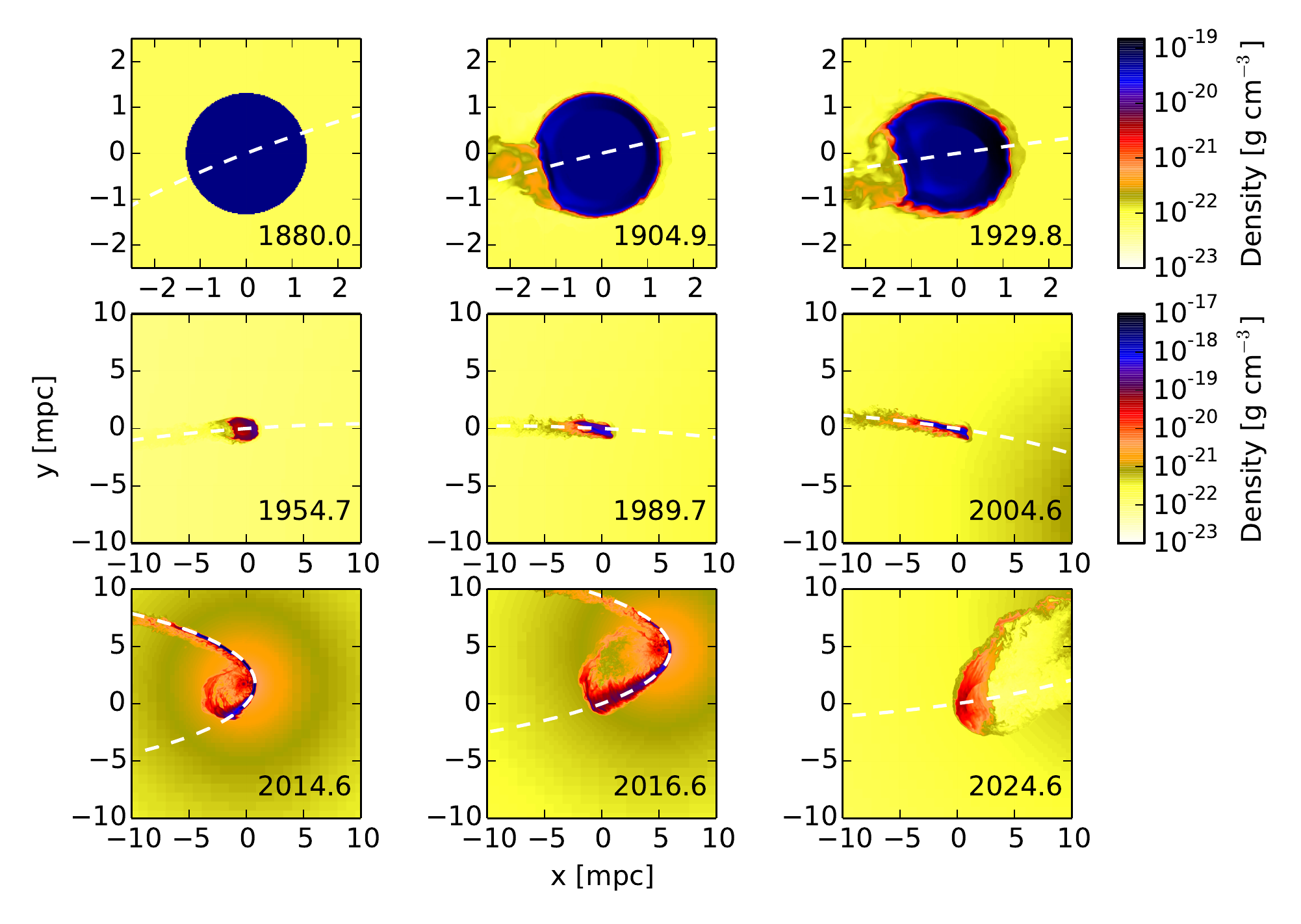}
\caption{Characteristic time snapshots of
  the evolution of the cloud in the standard model 1880M. Shown are cuts through
  the 3D density distribution within the orbital plane, centered on the
  nominal position along the orbit. Indicated is the simulation time
  of the respective snapshots. 
This time might be different with respect to the time when G2 is observed
at that position due to the time shift discussed in Sect.~\ref{sec:posvel_comp}.
The white dashed line refers to the nominal
  orbit. 
The cloud is first compressed due to the increasing
  pressure of the atmosphere with decreasing distance from the black
  hole (upper row), then suffers from hydrodynamical instabilities at
  its boundary (middle row) and gets tidally stretched along the orbit
  and partly accreted towards the central BH (lower row).
}
\label{fig:densevol_zoom}
\end{figure*}

The general evolution of the density distributions of the Compact Cloud models  
was already discussed in great detail in
\citet{Schartmann_12}. The 3D simulations show qualitatively the same
behavior. This is shown for model 1880M (see Table~\ref{tab:simparam}) in
Fig.~\ref{fig:densevol_zoom}.  We will refer to the latter simulation
as the {\it standard model} throughout the paper.
The clouds start with spherical shape and in pressure
  equilibrium with the atmosphere.
Moving closer to Sgr~A*, the surrounding density, temperature, and therefore
also the pressure increase steeply, leading to a slight spherical
contraction of the cloud (first row in Fig.~\ref{fig:densevol_zoom}). 
The dominant effect during the evolution
close to pericenter is the
tidal interaction with the BH, stretching the cloud in direction towards the
BH and squeezing it perpendicular to it
(Fig.\,\ref{fig:densevol_zoom}, middle row). 
Ram pressure interaction with the dense atmosphere
surrounding the BH leads to a compression of the front part of
the cloud. Additionally, the cloud experiences fluid instabilities
when interacting with the atmosphere, dominated by gas stripping
initiated by the Kelvin-Helmholtz instability.
The last row in Fig.~\ref{fig:densevol_zoom} shows the
    currently ongoing pericenter passage and the late time
    evolution. The interaction with the atmosphere then leads to the 
formation of a nozzle-like structure feeding gas towards the central
massive BH (Fig.~\ref{fig:densevol_zoom}, lower right panel).
A starting date as early as applied in some of the simulations
presented here will slightly
change the orbit. This is because of a slight decrease of the kinetic energy and
angular momentum redistribution due to the interaction with the
ambient atmosphere as well as the ram pressure compression of the
front part of the cloud (see discussion in \citealp{Burkert_12}). 
The biggest problem for a direct comparison with the recent
observations is that the positions are slightly shifted. 
The fitting of a new orbit taking hydrodynamical effects into account
is beyond the scope of this publication, but we account
for this effect by allowing for a time shift of our simulations relative to
the observations. 
In future, more detailed investigations of this shift in
orbital parameters compared to the purely ballistic orbit could 
provide valuable information on the structure of the surrounding 
diffuse gas component.

\section{Comparison to observations}
\label{sec:obscomp}

We describe G2's appearance on the sky -- based on our
simulations -- as observable with
the SINFONI instrument in Sect.~\ref{sec:brg_onsky}. The
comparison with the observations is split into two parts: (1)
position-velocity diagrams enable us to constrain the cloud's size
evolution and morphology and (2) the total Brackett-$\gamma$
light curve gives additional constraints for our models.

\subsection{G2's appearance on the sky}
\label{sec:brg_onsky}

\begin{figure}[b!]
\epsscale{1.2}
\plotone{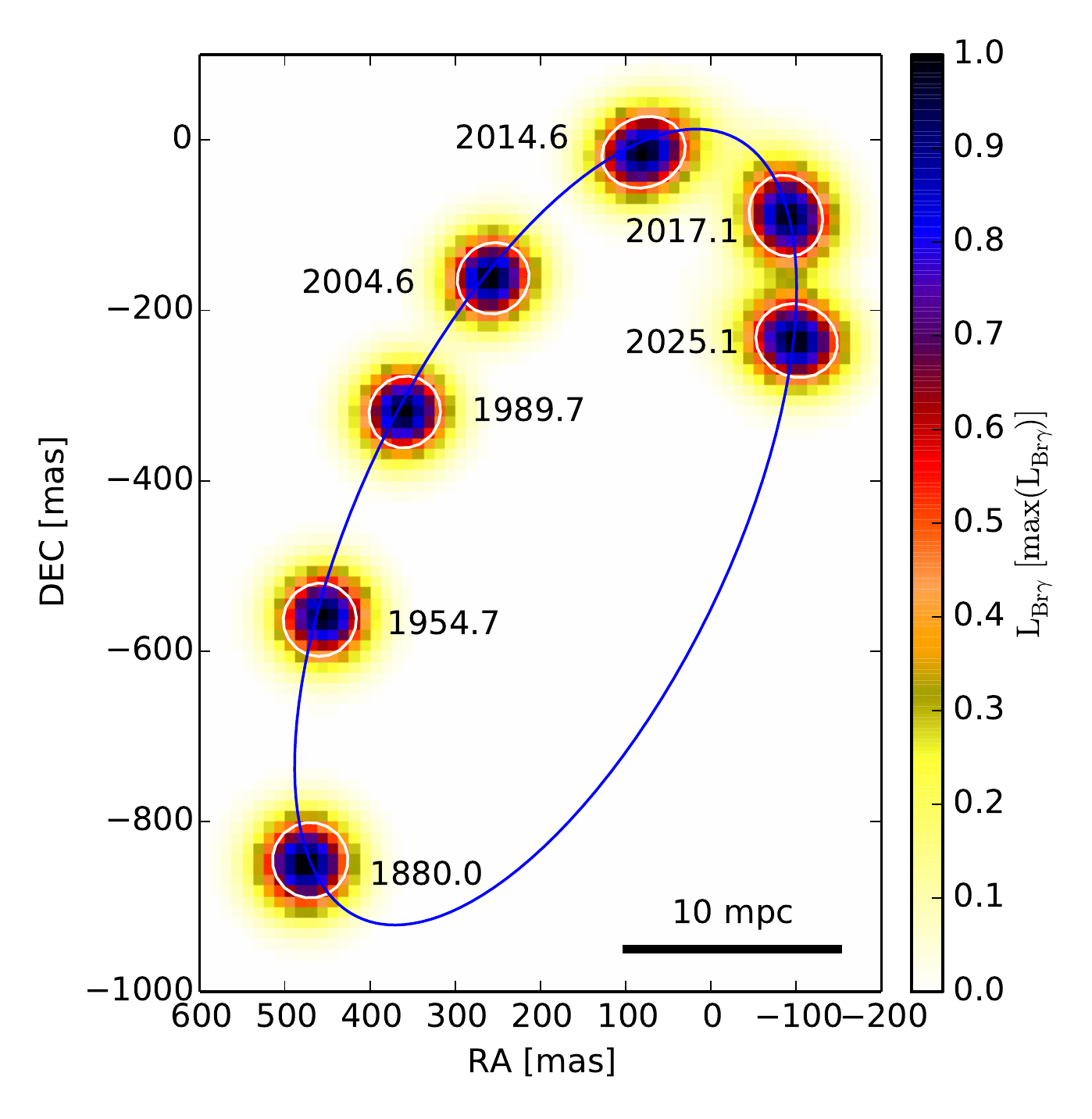}
\caption{Simulated, normalized  Brackett-$\gamma$ emission for the standard model 1880M projected on the
  sky and convolved with Gaussian distributions according to the instrumental FWHM
  values. Overlaid are white contours where half the maximum of the
  fitted 2D Gaussian is reached. Annotated is the
    simulation time. The cloud starts only marginally resolved and
    spherically symmetric. Moving closer to pericenter, it elongates
    in direction of motion and after pericenter in perpendicular direction.
}
\label{fig:brg_onsky}
\end{figure}

In order to be able to directly compare our simulations to the observed
Brackett-$\gamma$ emission maps and position-velocity diagrams,
we transfer our AMR data into a 3D data cube
spanned by right ascension (RA), declination (DEC) and line of sight
velocity ($v_{\mathrm{los}}$) with the same pixel sizes (12.5\,mas in
coordinate direction and 69.6\,km\,s$^{-1}$ in velocity direction) as
the one obtained from the SINFONI observations\footnote{A distance to
  the Galactic Center of 8.33\,kpc is assumed \citep{Gillessen_09,Ghez_08}.}. 
This projection makes use of the orbital elements as derived by
\citet[][Table~1, right column]{Gillessen_13b} for the {\it Brackett-$\gamma$ orbit}.  
The cells of the cube are filled with the respective total luminosity in the
Brackett-$\gamma$ line as derived with the formalism given in
\citet{Ballone_13}. In brief, we estimate the
Brackett-$\gamma$ emissivity, following Case B recombination theory:

\begin{eqnarray}
\label{equ:brg_emissivity}
j_{\mathrm{Br}\gamma} = 3.44\times10^{-27}\,\left( \frac{T}{10^4\,\mathrm{K}}\right)^{-1.09}\,n_{\mathrm{p}}\,n_{\mathrm{e}}\,\mathrm{erg\,s^{-1}\,cm^{3}},
\end{eqnarray}

where $T$ is the gas temperature and $n_{\mathrm{p}}$ and $n_{\mathrm{e}}$ are the proton and
electron number densities within the respective grid cell\footnote{Here we assume that the cloud is fully ionized. This however
sensitively depends on the assumed ionizing flux from the surrounding
stars, see discussion in A.~Ballone et al., 2015, in preparation.}.
Eq.~\ref{equ:brg_emissivity} is the result of a fit to the
recombination coefficients (for $T=5,000$, 10,000 and 20,000\,K)
as presented in \citet{Osterbrock_06}, which is in reasonable
agreement with the approximations found in \citet{Ferland_80} and \citet{Hamann_99}. 
To obtain the total Brackett-$\gamma$ luminosity within the mock
SINFONI cube, we integrate over the volume of the 
corresponding region of the simulation data, 
wherever our cloud tracer value
is above $10^{-4}$. Hence, the temperature change due to mixing
with the hot surrounding atmosphere is taken into 
account. 
The resulting cube is then smeared out
by applying a luminosity conserving Gaussian convolution with a full width at half maximum (FWHM) size of 81\,mas in
coordinate directions and 120\,km\,s$^{-1}$ along the velocity axis,
mimicking the characteristics of the SINFONI observations.
This convolution makes the PV diagrams shown in the following
independent of simulation resolution.

\begin{figure}
\epsscale{1.2}
\plotone{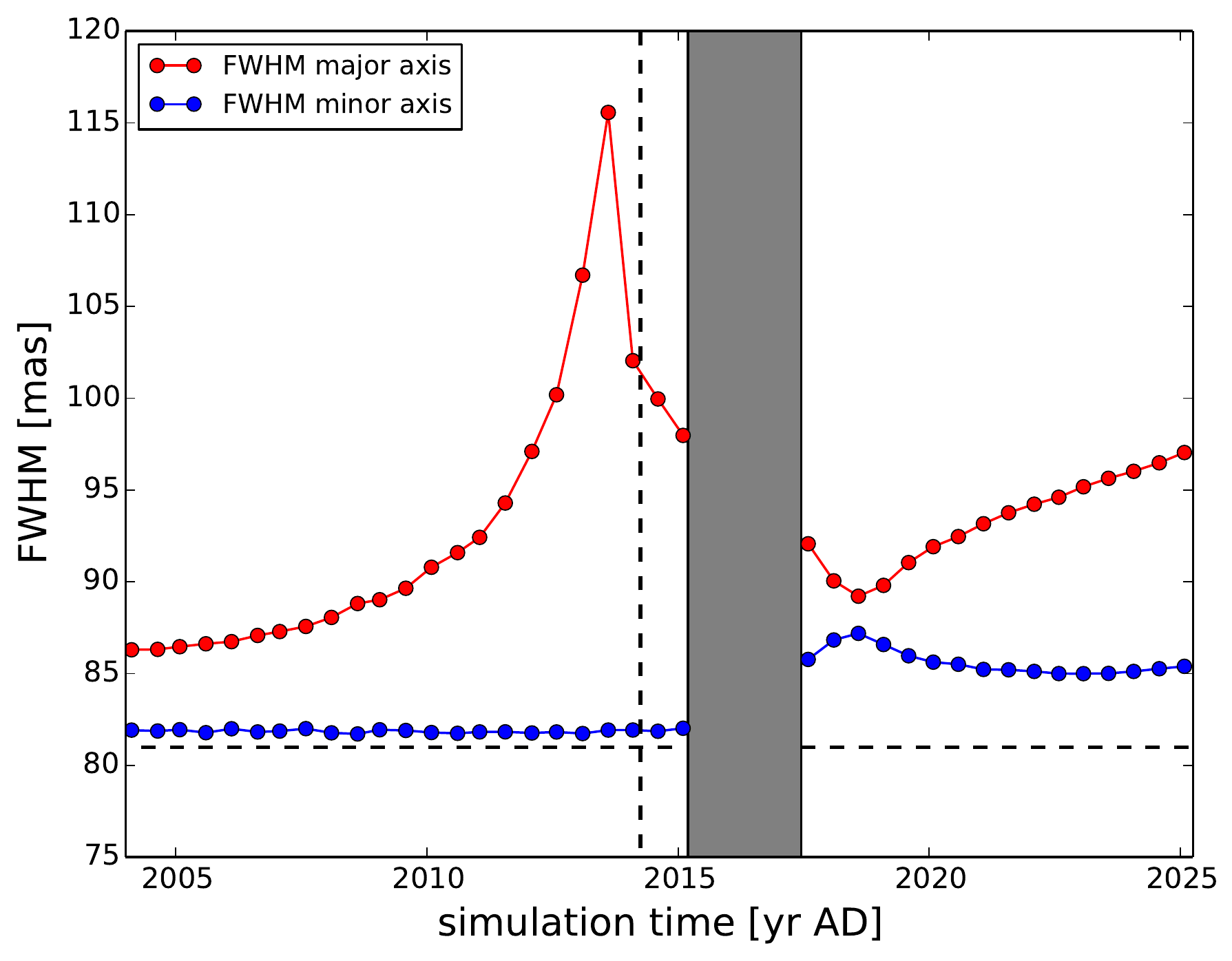}
\caption{Evolution of the full width at half maximum of
  the normalized Brackett-$\gamma$ emission projected on the
  sky for the standard model 1880M (compare to Fig.~\ref{fig:brg_onsky}). The black dashed horizontal line denotes 
  the instrumental FWHM of 81\,mas, with which the simulations 
  were convolved. The vertical dashed line marks the position of
  nominal pericenter passage. Gray-shaded is the time interval in
  which the Brackett-$\gamma$ emission of the cloud significantly
  deviates from a Gaussian shape.
  No time shift is applied to the simulation data in this
    figure. Clearly visible is the tidal stretching of the cloud before pericenter.
}
\label{fig:fwhm_brg_onsky}
\end{figure}

The resulting projection on the sky (neglecting foreground extinction) is shown in
Fig.~\ref{fig:brg_onsky}. We fit 2D Gaussians and overlay the contour at half
the maximum of the fitted 2D Gaussian for various timesteps as
indicated. 
The FWHM of the fitted Gaussian distributions along the major and
minor axis are shown as a function of time in Fig.~\ref{fig:fwhm_brg_onsky}. 
The cloud stays spherically symmetric until
roughly 1920, when it begins to slightly elongate in direction of
motion until shortly after pericenter. 
The reason for the elongation is the decrease of the minor axis towards the
instrumental FWHM of 81\,mas from roughly 1920 onwards due to the 
compression from the increasing atmospheric pressure, and later on, due to
tidal compression. The FWHM of
the major axis decreases 
only slightly, as ram pressure and atmospheric compression are partly
balanced by tidal stretching. 
From roughly 2000 onward the tidal forces dominate and
stretch the cloud significantly. Close to pericenter a FWHM of
roughly 115\,mas is reached. At this time the front part of
the cloud already starts to wrap around the BH. The tidal disruption
is faster closer to the BH and the cloud gas also heats up
significantly during pericenter passage (see discussion in
Sect.~\ref{sec:discussion}). 
These processes lead to a decrease of the fitted FWHM
close to pericenter and the maximum of the Brackett-$\gamma$ emission
also shifts backwards within the
cloud. During and shortly after pericenter passage, the cloud is not
well described by a Gaussian distribution (given by the gray-shaded
interval in Fig.~\ref{fig:fwhm_brg_onsky}). 
At the latest 
snapshots shown in Fig.~\ref{fig:brg_onsky} after the cloud has
started to contract again in direction
of motion (see Fig.~\ref{fig:densevol_zoom}), the
elongation changes to be perpendicular to the orbital motion.
This is similar to the morphology of the blue channel image representing
the Brackett-$\gamma$ emission of G1 in Fig.~6 of \citealp{Pfuhl_15}. 
Due to the 
uncertainties discussed earlier, however, the evolution past pericenter is
only partly physical in our simulations and has to be analyzed with
care, especially because of the steep drop of the
Brackett-$\gamma$ luminosity (Sect.~\ref{sec:brg_evol}).   
Even models with a very wide-stretched
gas density distribution close to pericenter
(e.~g.~model 1880M, see Fig.~\ref{fig:densevol_zoom}) result in only 
moderately extended Gaussian FWHM following our sky projection
procedure. Until the end of our simulation (at around
2025), we measure FWHM values similar to the ones before
pericenter. Therefore, our simulated cloud would be interpreted as
a rather compact object.

\subsection{Position-velocity diagrams}
\label{sec:posvel_comp}

Position-velocity (PV) diagrams give us the possibility
to gain observational insight into the dynamical evolution
and the tidal disruption of the gas cloud (see Fig.~\ref{fig:pvcomparison},
left panels). To this end, we
derive mock SINFONI data cubes from our simulation data (see
Sect.~\ref{sec:brg_onsky}) and 
extract the PV diagrams along the orbit from the data cube
in a similar way as done for the observations in order to allow
for a direct comparison:
To derive the relative position along the orbit for all
simulation grid points, we first project them on the orbit. 
In a further step, we extract the standard deviation $\sigma$ of the fluctuation spectrum of 
the observed PV diagrams within a region unaffected by the cloud or
streamer. In order to allow for a better comparison by
  eye, this is then used to overlay the according white noise onto
the simulated PV diagram. For the numerical calculation of the best
match, the simulated PV diagrams without noise are used. Both observed and simulated PV
diagrams are then normalized to the peak Brackett-$\gamma$ luminosity
of all pixels. This enables us to separate the constraints on the
size and stretching of the cloud from the total emission, and thereby
significantly simplifies finding a best-fit model. 
The simulated PV diagrams calculated in this way are
  shown in the middle column of Fig.~\ref{fig:pvcomparison}.
To allow for a quantitative evaluation of
the comparison with observations, we calculate normalized residuals
in a rectangular region with $21\times21$\footnote{The
  chosen size ensures that the cloud is fully within the region for
  all epochs and excludes emission from G2's tail or G1.} pixels surrounding the
maximum within the G2 cloud in the observed PV diagram in the
following way: 

\begin{eqnarray}
\label{equ:qualmatch}
\delta^2 (v_{\mathrm{los}},x_{\mathrm{orb}}) =
  \left(\frac{PV_{\mathrm{obs}}-PV_{\mathrm{sim}}}{\sigma}\right)^2,
\end{eqnarray}
where all arrays are functions of the line of sight velocity ($v_{\mathrm{los}}$) and
the relative position along the orbit ($x_{\mathrm{orb}}$) and
$PV_\mathrm{sim}$ refers to the normalized simulated PV diagram
without noise.
The resulting $\delta^2$-arrays are shown in the right hand panels of
Fig.~\ref{fig:pvcomparison}. 
Summing up of the two-dimensional $\delta^2$ array leads
us to the $\chi_\mathrm{r}^2$ value
which we use to characterize the quality of the match:

\begin{eqnarray}
  \label{equ:chi2}
 \chi_\mathrm{r}^2 = \frac{1}{N-1}\,\sqrt{\sum \delta^2},
\end{eqnarray}

where $N=421$ is the total number of pixels of the $\delta^2$
array.
The closer $\chi_\mathrm{r}^2$ to one, the better the
match with the data. In this metric, a model with a PV diagram equal
to zero everywhere results in $\chi_\mathrm{r}^2$ values of 4.37
for the 2008 observation, 7.93 for 2010, 13.25 for 2011, 14.53 for
2012 and 12.78 for 2013.
This procedure is repeated for all simulation snapshots with a time
interval of typically 0.5\,yr, which we compare to all observational
epochs in this work. For a more detailed description of the construction of PV
diagrams from our AMR data we refer to A.~Ballone et al. (2015, in preparation).

\subsubsection{The standard model}

\begin{figure*}
\epsscale{1.0}
\plotone{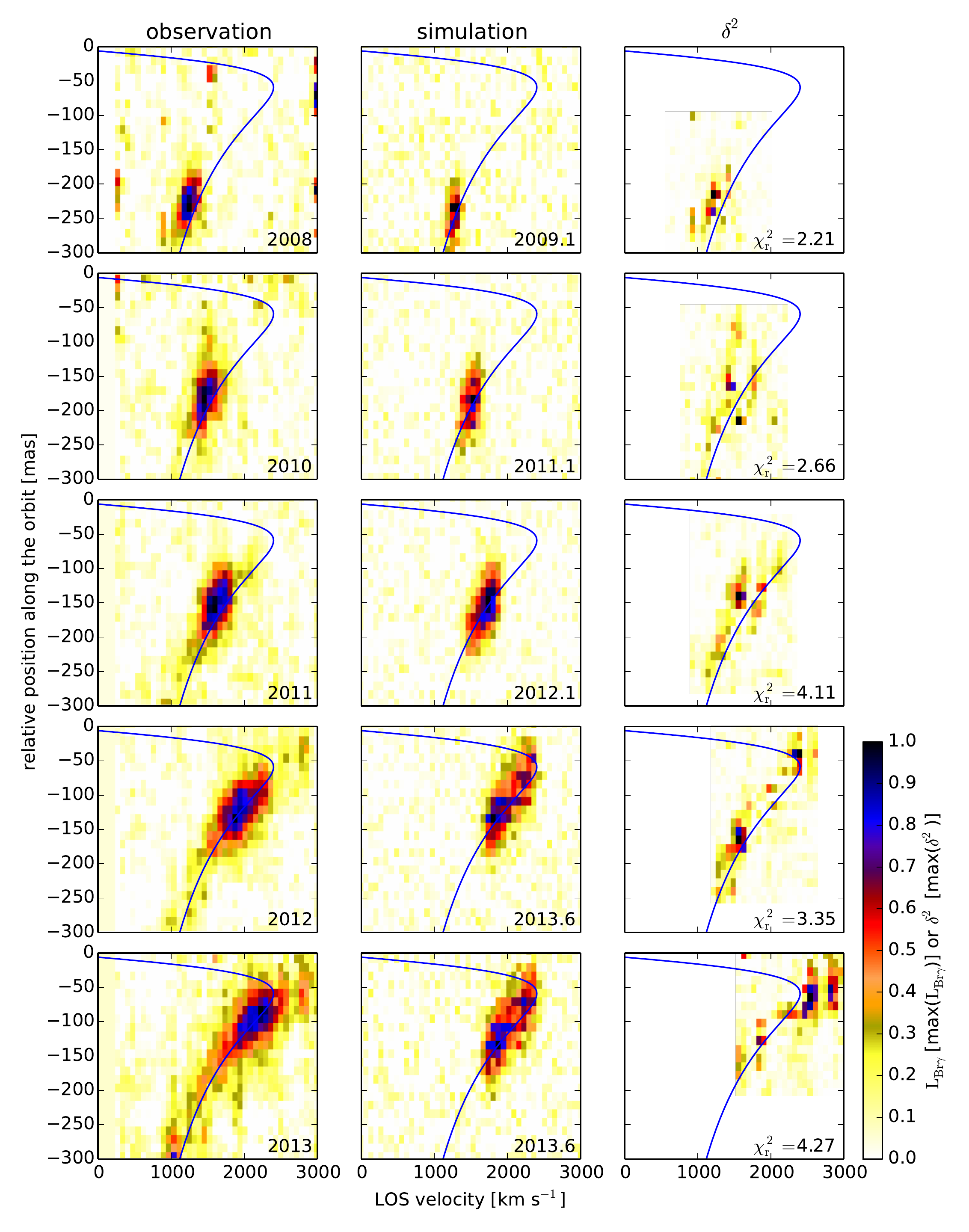}
\caption{Comparison of observed (left panels) and best-fit simulated (middle panels) position-velocity
  diagrams of our standard model 1880M for several observational
  epochs as indicated in the left panels. Indicated in the middle
  panel is the simulation time. The simulation data has been
  convolved with a Gaussian and noise has been added (see
  Sect.~\ref{sec:posvel_comp}). 
  The right hand panels show the normalized residuals
  $\delta^2$ (given in Eq.\,\ref{equ:qualmatch}). These are based
  on the simulated PV diagrams without added noise.
  The resulting $\chi_\mathrm{r}^2$ values are given in the lower
  right corners. 
All
  diagrams are normalized to their respective maxima. The
  observational data is adapted from
  \citet{Gillessen_12,Gillessen_13a,Gillessen_13b} and
  \citet{Pfuhl_15}.
A good match with observations is found, except for the high velocity
gas at small distances from pericenter in 2013.
}
\label{fig:pvcomparison}
\end{figure*}

The result of this procedure is summarized in
Fig. ~\ref{fig:pvcomparison} for the standard model. 
For selected observation epochs (various
rows), the left hand side panel displays the
observed PV diagram, the middle one is the best-fitting simulation
snapshot and the right hand side panel displays the normalized
residual $\delta^2$ array. 
The years 2004 and 2006 have too low signal-to-noise ratios in order 
to allow for a meaningful comparison and are omitted here.
The best match is not obtained at
the simulation time of the observed epoch, but with roughly one year
delay. 
The exception of the 2013 epoch will be discussed
in more detail below and in Sect.~\ref{sec:start_time_study}. 
This is a small effect given the
orbital time of roughly $391 \pm 66$ years,
but needs to be taken into account for a thorough comparison with
observations. 
The reason for this time shift is the hydrodynamical
interaction of the cloud with the hot atmosphere. Due to the early
starting time of the simulation, the ram pressure
interaction has accumulated and becomes noticeable in form of a deceleration of the cloud,
which thereby ends up on a slightly different orbit. This change in
orbital evolution is approximated by the introduction of the mentioned
time shift.   
The 2013 epoch is a special case, as by then, the front part of the cloud was already in
the process of pericenter passage. 
Due to the tidally stretched appearance of the cloud,
the pericenter passage takes up to several years
when taking low density gas into account. 
Given the remaining problems
in our simulations during pericenter passage (see discussion in
Sect.~\ref{sec:discussion}), we decided to restrict the quantitative
part of our analysis to the pre pericenter evolution. The middle 
panels of the 2012 and 2013 epochs also demonstrate the influence of
the added noise to the appearance of the cloud in the PV diagrams, 
as they are showing the same simulation snapshots.

\subsubsection{Starting time study}
\label{sec:start_time_study}

\begin{figure}
\epsscale{1.2}
\plotone{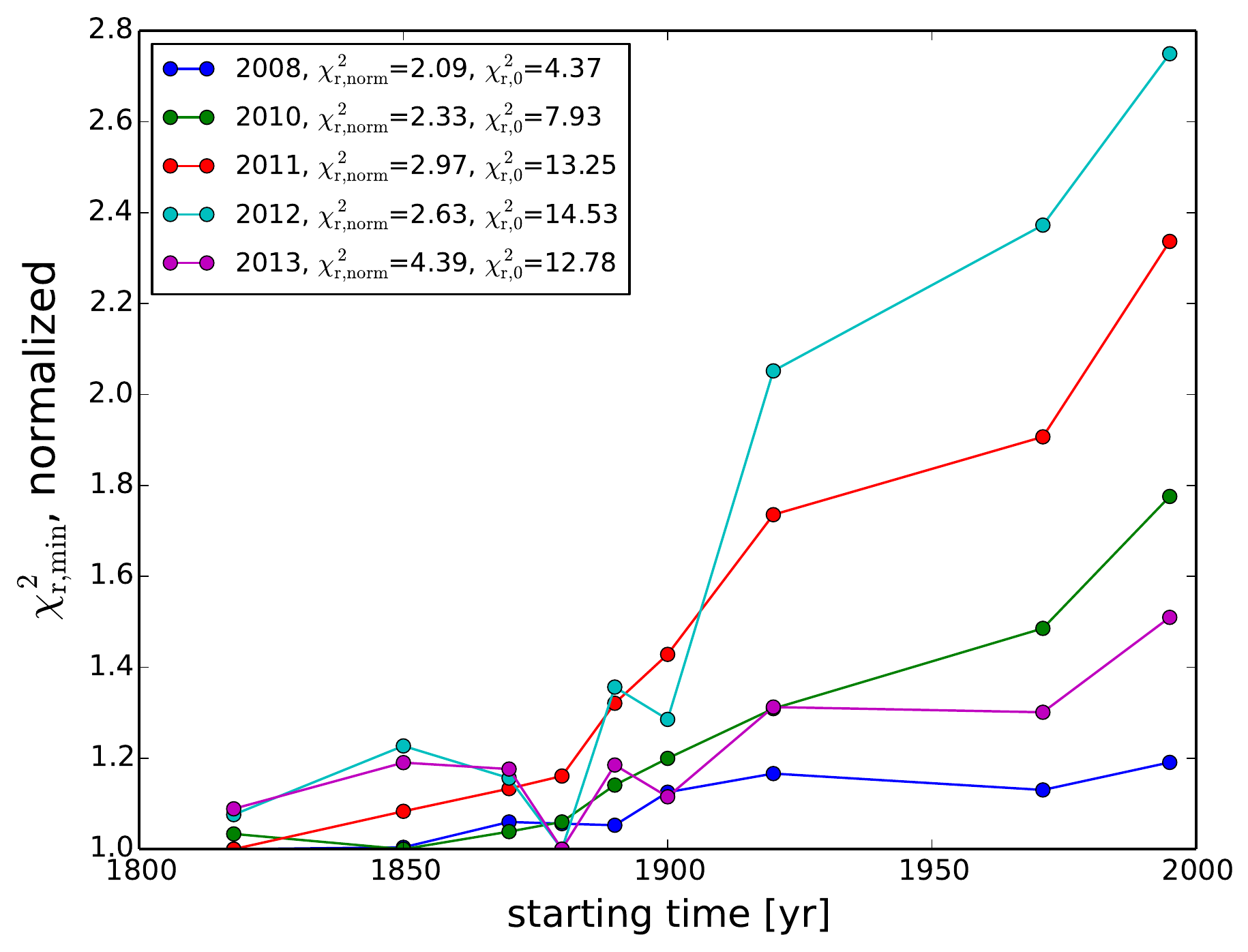}
\caption{Normalized $\chi_\mathrm{r,min}^2$-values (see Eq.~\ref{equ:chi2}) for all starting times of the
  simulations, normalized to their respective minima (as
    given in the legend by $\chi_\mathrm{r,norm}^2$) and color-coded
  according to the observation epoch. The low resolution simulations
  have been used for this analysis. 
  $\chi_\mathrm{r,0}^2$ corresponds to a PV
  diagram filled with zeros.
Compare to Fig.~\ref{fig:min_timeshifts_obsepoch_cut} for the
  applied time shifts and Table~\ref{tab:rchi2vals} for absolute values
  of $\chi_\mathrm{r}^2$.
This indicates that the closest match with observed PV diagrams is
reached for starting times close to or earlier than 1900.
}
\label{fig:min_rchi2_starting_time_norm}
\end{figure}

\begin{figure}
\epsscale{1.2}
\plotone{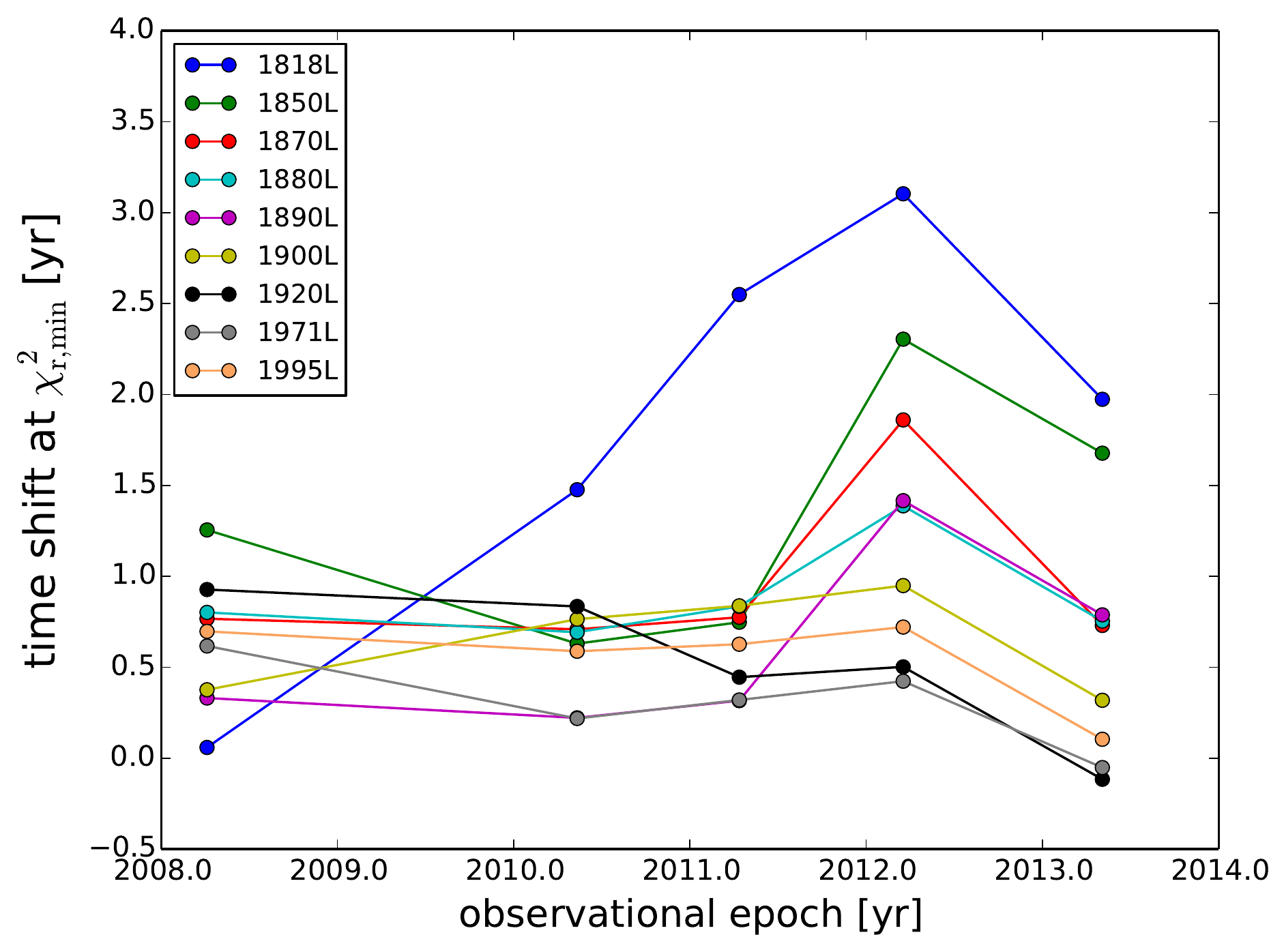}
\caption{Resulting time shifts from the comparison of simulated to
  observed PV diagrams as described
  in Sect.~\ref{sec:posvel_comp}. Compare to
  Fig.~\ref{fig:min_rchi2_starting_time_norm} and Table~\ref{tab:rchi2vals} for the corresponding
  $\chi_\mathrm{r,min}^2$ parameters. 
 The tentative increase towards
  later epochs is due to the interaction with the atmosphere. The drop
  for the 2013 epoch is caused by the fact that part of the cloud has
  passed pericenter and we are only comparing red-shifted emission.
}
\label{fig:min_timeshifts_obsepoch_cut}
\end{figure}

Fig.~\ref{fig:min_rchi2_starting_time_norm} displays a summary of this
analysis for a number of simulations (see Table~\ref{tab:simparam}) 
in which we change the starting time of the cloud in
pressure equilibrium for the five observing epochs
2008, 2010, 2011, 2012 and 2013. 
The $\chi_\mathrm{r}^2$ parameters shown are the
minima after comparing all simulation snapshots with the respective
observation. Each curve corresponding to an observational
epoch is normalised to its minimum value, which are given in the
legend of the plot. Absolute $\chi_\mathrm{r}^2$ are given for some
simulations in Table~\ref{tab:rchi2vals}. 
The different slopes of the curves for various
  observational epochs 
  relate to the increasing offset of the observed PV diagram from the
  nominal orbit, which can not be accounted for in the
  simulations and the different noise levels. The best constraint for our simulations is given by the
  2011 and 2012 epochs (showing the steepest curves) and almost no
  constraint can be derived from the 2008 epoch (flat distribution). 
The resulting time shifts with respect to the simulation
time are up to 3 years and tend to increase towards later
observing epochs (Fig.~\ref{fig:min_timeshifts_obsepoch_cut}).
Given that a range of 80 years in starting time lead to 
a good comparison with the data (given by the plateau in
Fig.~\ref{fig:min_rchi2_starting_time_norm} for starting times earlier
than approximately 1900), it is a small effect.
This shows that deviations from a ballistic orbit are minor in this
evolutionary stage.
The increase towards later observing epochs might be evidence for a too strong
  interaction of the cloud with the atmosphere, which could either be
  caused by a too large cross section of the cloud in this
  evolutionary phase or necessitate a change of the structure of the
  assumed atmosphere in the corresponding radius regime. 
  However, part of the increase could also be related
  to different data quality of the epochs, the uncertainty of the
  orbit and the (partly artificial) mixing of cloud material with the 
  atmosphere (see Sect.~\ref{sec:mixing_plateau}). 
For the 2013 observation epoch the time shifts
decrease again, which is consistent with the observational finding
that part of the cloud has already passed pericenter at that time \citep{Gillessen_13b}.
As a consequence, the best comparison with the observations in
2012 and 2013 is reached for the same simulation snapshot (2013.6), as
can be seen in the last two rows of Fig.~\ref{fig:pvcomparison}.

\begin{figure*}
\epsscale{1.0}
\plotone{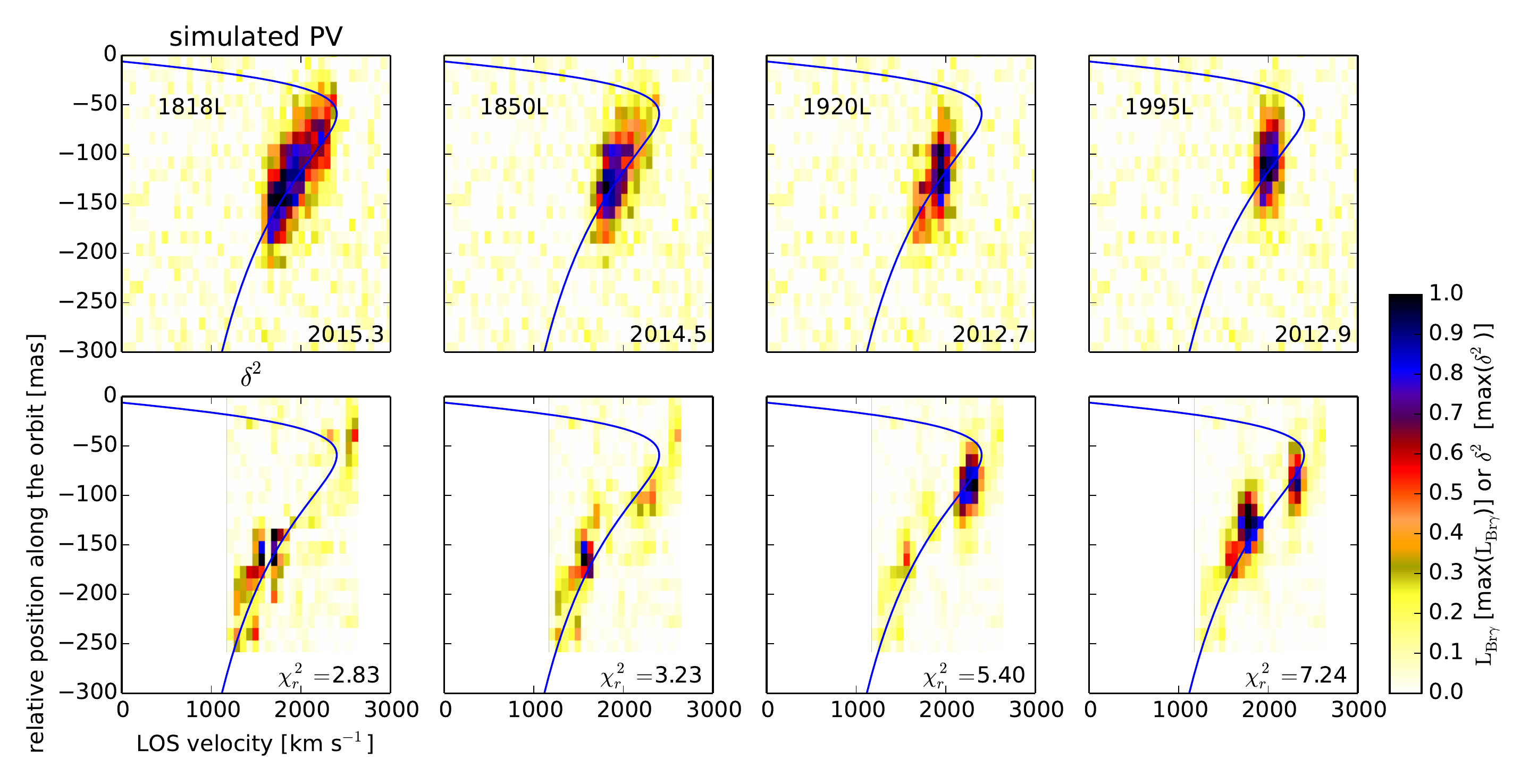}
\caption{Comparison of simulated position-velocity diagrams 
(including
  noise extracted from the observed ones) with the
observations in 2012 for simulations
with various starting times as annotated (upper row). The snapshot simulation time
is given in the lower right corner. The lower row shows the
corresponding normalized residuals arrays $\delta^2$ (see description in Sect.~\ref{sec:posvel_comp}). The blue
curve denotes the nominal orbit of G2 as derived from observations in
\citet{Gillessen_13b}. 
The plot clearly demonstrates that an earlier
starting point results in a better match with observations.
}
\label{fig:pvcomparison_somesim}
\end{figure*}

Fig.~\ref{fig:pvcomparison_somesim} shows the comparison of the
position-velocity diagrams of a subset of the
simulations for the observing epoch of 2012. The upper row
shows the simulated PV diagrams including noise and the lower row depicts the
corresponding normalized residuals $\delta^2$. 
The residuals indicate that the models with the late
  starting times (1920L and 1995L in this example) underestimate the
  size of the cloud and a good match is reached for the early starting
  times. This is also quantified in
  Table~\ref{tab:rchi2vals}. E.~g.~for the 2012 observation, the
  standard model
  reaches a factor of 2.2 lower $\chi_\mathrm{r}^2$ value compared to the 1995L
  model and a factor of 4.3 lower compared to a PV diagram with zeros
  everywhere.
Altogether, this clearly shows that an early starting
time at around the year 1900 or earlier is favoured by this detailed comparison
to observations. The reason for this is the tidal stretching which
needs to fully unfold in the simulations and finally leads to the observed structure of
the cloud. The plateau in
Fig.~\ref{fig:min_rchi2_starting_time_norm} at early starting times is
caused by the slow evolution of the cloud close to apo center.

\begin{table}
\begin{center}
\caption{Resulting $\chi_\mathrm{r}^2$ values (Eq.~\ref{equ:chi2}) for the comparison
    of simulated and observed PV diagrams for selected models.\label{tab:rchi2vals}}
\begin{tabular}{lcrrrrrrrr}
\tableline\tableline
name & 2008 & 2010 & 2011 & 2012 & 2013\\
\tableline
1880M & 2.21 & 2.66 & 4.11 & 3.35 & 4.27\\
1995L & 2.49 & 4.13 & 6.94 & 7.24 & 6.62\\
PV=0  & 4.37 & 7.93 & 13.25 & 14.53 & 12.78\\
\tableline
\end{tabular}
\tablecomments{PV=0 refers to the extreme case of a simulated PV diagram equal to zero everywhere.}
\end{center}
\end{table}

\subsection{The Brackett-$\gamma$ luminosity evolution}
\label{sec:brg_evol}

\begin{figure}
\epsscale{1.2}
\plotone{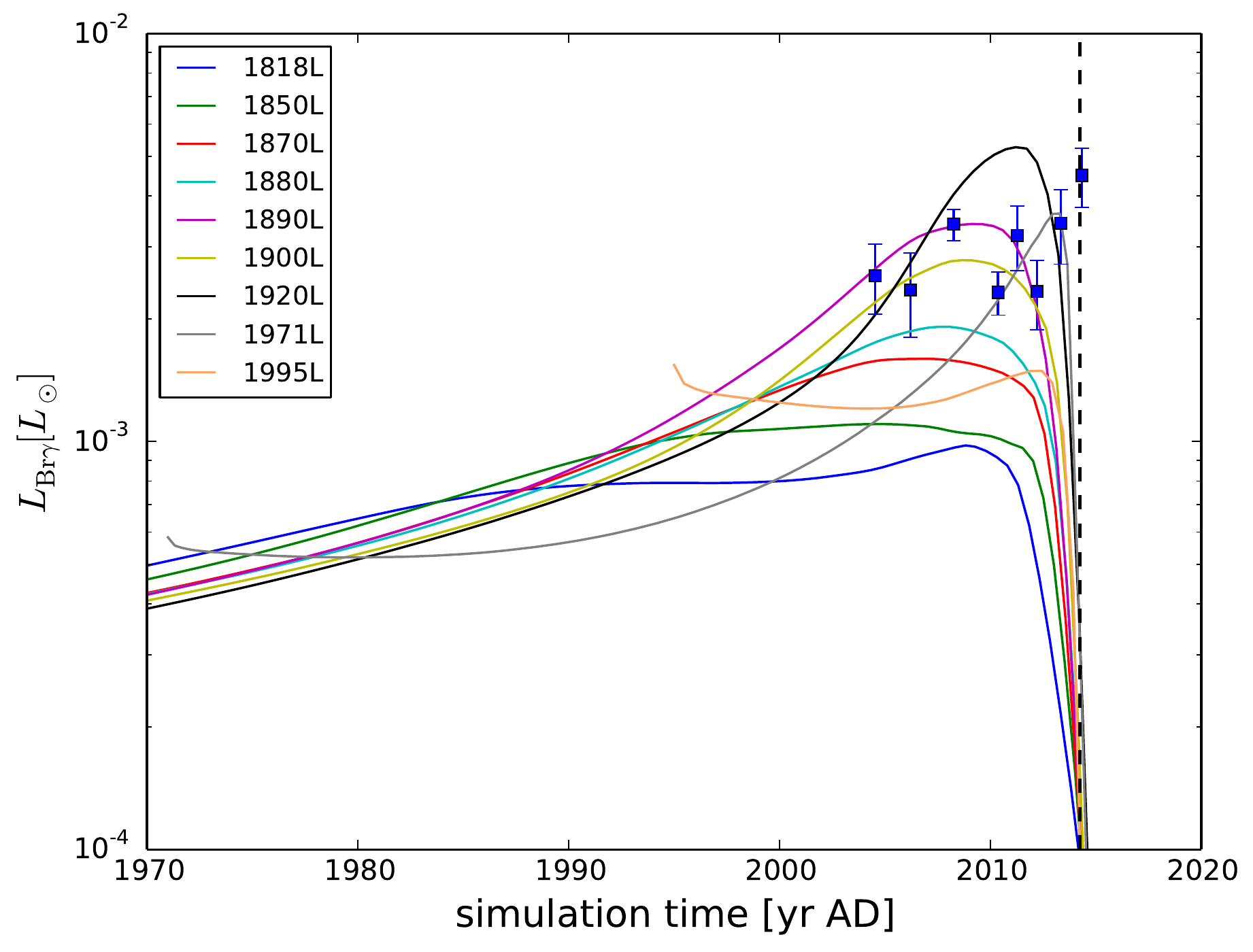}
\caption{Total Brackett-$\gamma$ luminosity evolution for simulations
  with various starting times as indicated in the legend. Overplotted
  as blue symbols with error bars
  are the observational data from
  \citet{Gillessen_12,Gillessen_13a,Gillessen_13b} and
  \citet{Pfuhl_15}.
    No time shift is applied to the simulation data in this
    figure. The early rising signature due to atmospheric compression
    is followed by a plateau due to mixing with the atmosphere and a
    steep drop, which is caused by the strong expansion of the cloud
    post pericenter. The resolution dependence is shown in Fig.~\ref{fig:brglum_resolution}. 
}
\label{fig:brglum_startingtime}
\end{figure}

In order to derive the time evolution of the total Brackett-$\gamma$
emission (see Fig.~\ref{fig:brglum_startingtime}), we use Eq.~\ref{equ:brg_emissivity} and the formalism
described in Sect.~\ref{sec:brg_onsky}.
Because of the homogeneous density and temperature
distribution of the initial clouds and the gradient in atmospheric pressure, only the central
part of  the cloud is in pressure equilibrium and the very first phase
of the evolution is given by this slight pressure adjustment. Given the power-law
assumption for the atmospheric profiles, this radial pressure
inequality is stronger
when starting the cloud closer to Sgr~A*. 
This is directly visible in
the initial drop of the Brackett-$\gamma$ luminosity close to the
starting point of the simulations
(Fig.~\ref{fig:brglum_startingtime}) when comparing the 1995 light curve
with the 1971 one.

\subsubsection{Atmospheric compression}

The longest phase of the evolution of the
light curve is given by compression due to the pressure increase when
moving towards the central part of the atmosphere. The low sound
speed of approximately $c_\mathrm{s}=10\,\mathrm{km}\,\mathrm{s}^{-1}$ 
within the cloud with an initial sound crossing time of the
order of 50 to 100 years for the simulations shown here (see
Table~\ref{tab:simparam}) leads to a slowly inward growing
spherical density enhancement and a slow compression of the cloud (see
Fig.~\ref{fig:densevol_zoom}, upper row). 
Despite having a constant mass and lower density at the
beginning, models with an earlier starting time end up with a higher Brackett-$\gamma$ luminosity
compared to the later starting simulations. The reason is
the formation of this dense, outer shell due to this atmospheric
pressure confinement and the scaling of the Brackett-$\gamma$
emissivity with $\rho^2$ (Eq.~\ref{equ:brg_emissivity}). This behaviour is visible in
between roughly 1950 and 1970 and hence does not show up in
Fig.~\ref{fig:brglum_startingtime}.   

\subsubsection{Mixing plateau}
\label{sec:mixing_plateau}

\begin{figure*}
\epsscale{1.0}
\plotone{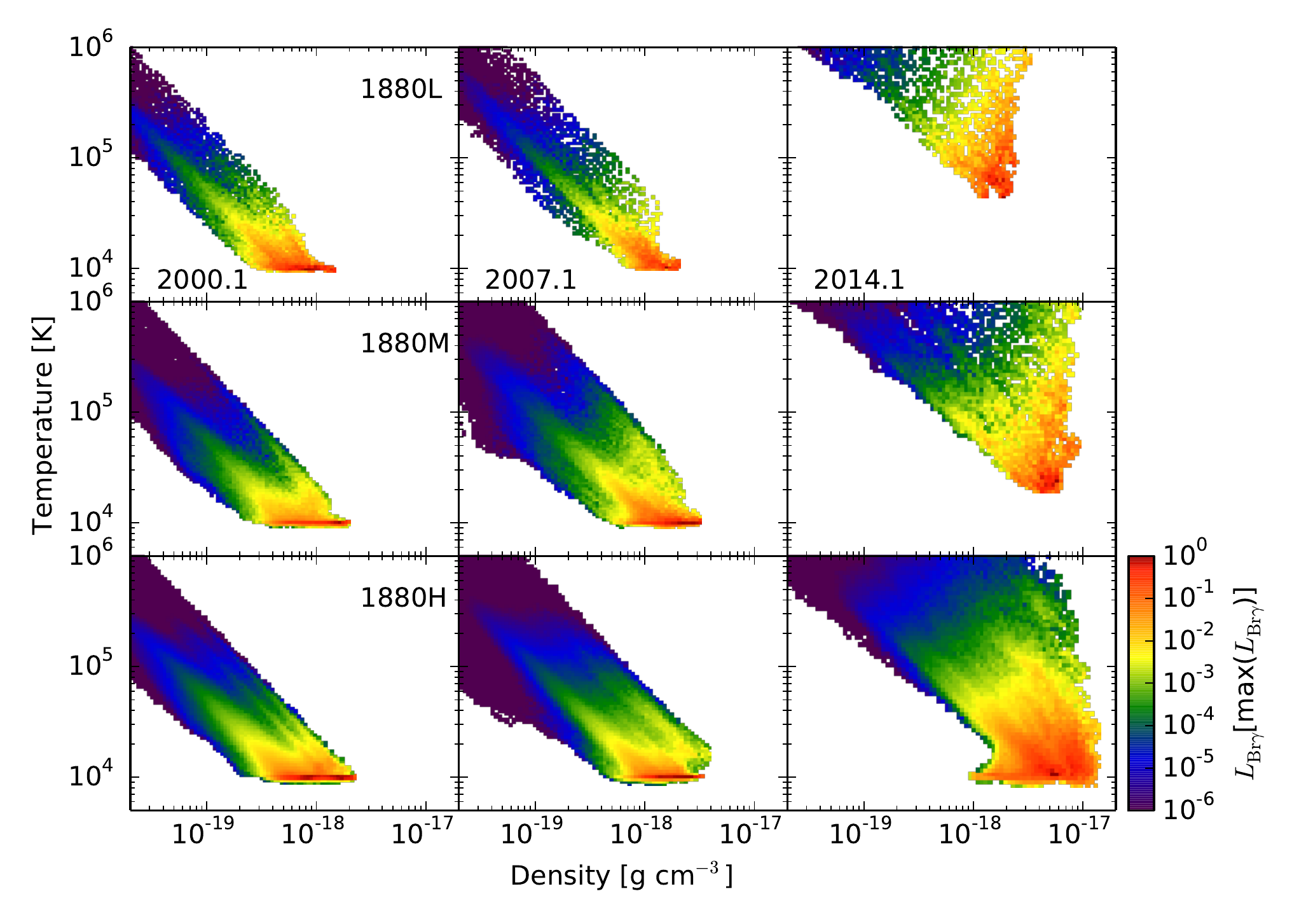}
\caption{Histograms of the Brackett-$\gamma$ emission as a function of
  the density and temperature for a starting time of the cloud in
  1880. Shown are models 1880L (upper row), 1880M (middle row) and
  1880H (lower row). The simulation name is made up of the starting
    time and a letter indicating the resolution, see
    Tab.~\ref{tab:simparam}. The histograms demonstrate the effect of
    the resolution and mixing of
the simulations on the Brackett-$\gamma$ emission.}
\label{fig:phaseplots}
\end{figure*}

The trend is only broken due to the limited resolution of the
simulations, which enhances the mixing with the
hot atmosphere and stalls further compression due to the
  atmospheric confinement. 
The mixing is partly physical and partly of 
numerical origin and has two effects. It (1) lowers the density of the cloud and
(2) leads to an increase of the  temperature, as is seen in the phase
diagrams for the Brackett-$\gamma$ emission presented in
Fig.~\ref{fig:phaseplots}. 
In these diagrams for the 1880 models the cloud starts in a single point with
  a temperature of $10^4$\,K and a density of $6.1\times
  10^{-20}\,\mathrm{g\,cm}^{-3}$. The
  atmospheric compression and the tidal evolution only change the
  density structure in these isothermal simulations, visible in the horizontal
  stretching of the red
  linear feature. The mixing with the atmosphere at the boundary of
  the cloud lowers the contribution of cloud material to the total
  density of the respective grid cells. The increasing contribution of the
  atmospheric gas heats up the cells. This dilution increases with
  distance from the cloud and leads to the apparent correlation of
  density and temperature in Fig. ~\ref{fig:phaseplots}. 
Both effects lower the
Brackett-$\gamma$ emission in our formalism (see
Eq.~\ref{equ:brg_emissivity}) and 
finally lead to the partial dissolution of the cloud. 

\begin{figure}
\epsscale{1.2}
\plotone{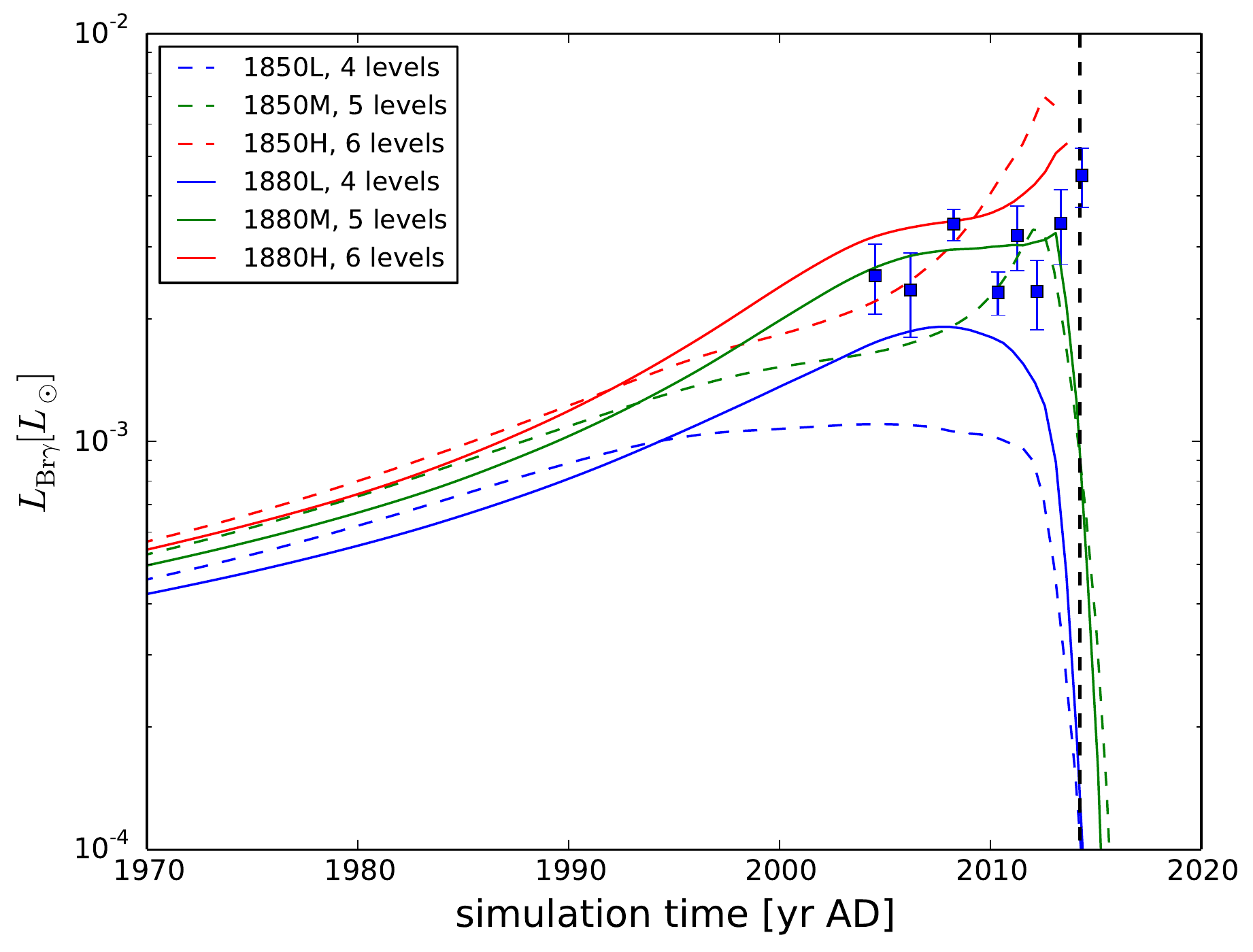}
\caption{Brackett-$\gamma$ luminosity evolution for two resolution
  studies. Overplotted
  as blue symbols with error bars
  are the observational data from
  \citet{Gillessen_12,Gillessen_13a,Gillessen_13b} and
  \citet{Pfuhl_15}.
No full convergence is reached. Only very high resolution simulations are able to resolve
the steep rise towards pericenter which is caused by tidal compression.
}
\label{fig:brglum_resolution}
\end{figure}

The effect of resolution on the light curves is shown in
Fig.~\ref{fig:brglum_resolution} for the 1850 and 1880 models. The lower the resolution, the
earlier the curves level off, 
as the numerical mixing is stronger due to the
coarser grid. The visible correlation with the
starting time of the simulation results from the longer time period in
which they are able to mix with the atmosphere. 
Many of the simulations show a well-defined, constant luminosity ``mixing plateau'', which
roughly follows the phase dominated by pressure confinement.

\subsubsection{Tidal compression}

The steep increase of the light curves later on is caused by tidal
compression. This increase is only present in light curves of well resolved
simulations. Under the assumption of hydrostatic equilibrium between tidal forces and
thermal pressure within the cloud  a Gaussian 
density distribution perpendicular to the momentum direction results. 
Its full width at half maximum is given by

\begin{eqnarray}
  \mathrm{FWHM} = 2\,\sqrt{2\,\ln{2}} \,\sqrt{\frac{r^3\,k_{\mathrm{B}}\,T}{G\,M_{\mathrm{BH}}\,\mu\,m_{\mathrm{u}}}},
\end{eqnarray}

where $r$ is the distance to Sgr~A*, $k_{\mathrm{B}}$ is the Boltzmann constant, $T$ is the temperature within the
cloud, $M_{\mathrm{BH}}$ is the mass of the central BH and
$\mu\,m_{\mathrm{u}}$ is the mean molecular weight (in grams) of the 
gas in the cloud. 

\begin{figure}
\epsscale{1.2}
\plotone{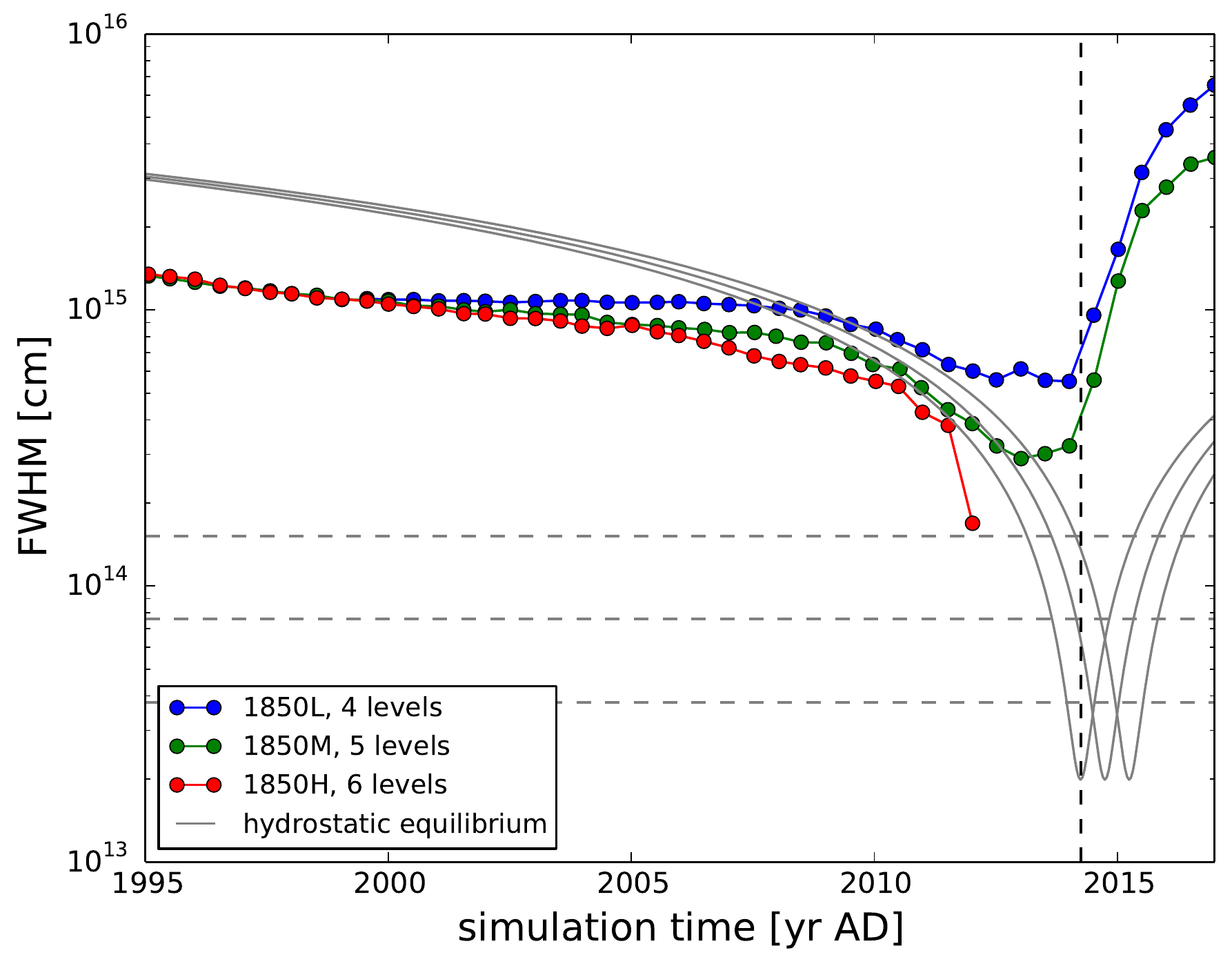}
\caption{FWHM of a Gaussian fit to the cloud density distribution perpendicular to the velocity vector
  at the position of the nominal orbit and surrounding
  it for simulations 1850L (blue), 1850M (green) and 1850H
  (red). The simulation name is made up of the starting
  time and a letter indicating the resolution, see Tab.~\ref{tab:simparam}.
  The gray curves refer to a
  simple model where hydrostatic equilibrium with the tidal
  compressional forces is assumed. The dashed horizontal lines show the resolution of the
  simulations and the dashed vertical line denotes the nominal
  pericenter. The limited resolution of the simulations prevents
  stronger compression and only the high resolution simulations result in a
  reasonable adaptation of the analytic model. Shown here are the
  results for a starting time in 1850,
  as this resolution effect is more pronounced for an earlier starting
  time.
}
\label{fig:dens_fwhm}
\end{figure}

To understand the late time behaviour better, we extract the
density distribution of the cloud perpendicular to the momentum vector
and fit Gaussian distributions. 
Fig.~\ref{fig:dens_fwhm} shows the full width at half maximum of an average
Gaussian distribution derived at ten equidistant positions on the
orbit with a maximum separation of $2.5\times10^{14}\,$cm along the orbit with respect to
the nominal position of G2 at the respective time.
An average is taken as the cloud has developed
  substructures at that time.
At the beginning of the simulations and for the longest part of their
evolution, the pressure forces due to the surrounding atmosphere
dominate over the tidal compressive forces, leading to FWHM values
below the ones derived for the simple analytic tidal disruption model.
This implies that only extremely high resolution simulations are able to resolve the
expected Gaussian density distribution close to pericenter passage.

\subsubsection{Post pericenter evolution}

After G2's pericenter passage, all our simulations predict a steep drop of the
Brackett-$\gamma$ luminosity (Fig.~\ref{fig:brglum_startingtime}), 
which is caused by heating due to mixing with the atmosphere
and the fast expansion of the cloud, which is only partially physical.
This is in tension with the relatively high level of blue-shifted
Brackett-$\gamma$ flux observed in the latest epoch \citep{Pfuhl_15}. 
The post-pericenter evolution is strongly
dependent on the assumed physics and a more detailed investigation will
be necessary.

\subsection{Possible interpretations}

The best comparison of the PV-diagrams with the observations
was reached for a starting time in 1900 and earlier. The
  PV diagrams give us the best constraints for our models, independent
of the resolution. 
A similar conclusion concerning the best starting time
  results from the comparison with the Brackett-$\gamma$
  luminosity evolution shown here. 
However, the luminosity evolution is not fully converged with
resolution for the simulations presented here and, therefore, has only
little constraining power.
In addition, the following needs to be considered:

\begin{enumerate}
\item Due to the sparse observational coverage, our choice of the atmosphere should be considered an
  assumption. E.~g.~changing the slope or the normalisation of the density
  distribution will change both the comparison to the PV diagrams
  and the evolution of the light curve.
\item The cloud mass is fixed to the observationally
    estimated value. Changing the initial cloud mass affects the size
    and luminosity evolution.
\item A dynamically relevant magnetic field strength has recently been
  found in the Galactic Center region by \citet{Eatough_13}. Depending on the exact
  strengths and morphologies of the field lines, this changes the mixing
  behaviour  \cite[e.~g.~][]{McCourt_15a} and additional, counterbalancing pressure forces might lead to an
  ``arresting'' of the cloud \citep{Shcherbakov_14, Schartmann_12}. Depending on the strength, the process
  might be able to counterbalance the atmospheric pressure, but only partially
  the tidal compression forces close to the nucleus, which would lead
  to a shape of the light curve similar to the observations with a
  plateau followed by a steep rise. 
\item A physical mixing process together with tidal compressive forces
  might be able to explain the currently observed luminosity evolution.
\end{enumerate}

\section{Discussion}
\label{sec:discussion}

The simulations of the Compact Cloud Scenario presented in this publication together with a detailed
comparison to available observations show that a starting point of the
cloud within the disk of young stars is favored by these models.

This is in tension with our previous analysis in \citet[][Fig.~2]{Schartmann_12}
which came to the conclusion that a starting point in 1995 yields the
best comparison with the observations. The main reason is that the latter was based
on a simple test particle analysis, not taking the internal structure
of the cloud into account (and the choice of a somewhat arbitrary
contour line for this very first comparison). Only 3D hydrodynamical simulations allow
such a detailed comparison as presented here. 
In this updated analysis we find that the cloud needs to be larger
than previously thought (see Fig.~\ref{fig:pvcomparison_somesim}). This is reached by starting the cloud
earlier on the orbit, leading to a larger initial size as well as a
stronger tidal stretching. 
One should also keep in
mind that the best-fit models derived here represent 
only one specific solution for the assumed mass of
the cloud and choice of parameters for our idealized atmosphere. 
Given the low number of observational constraints for the atmosphere,
especially in the region where the cloud is observed, large
uncertainties exist concerning density and temperature structure.
We also know that the actual atmosphere has a
significant inflow and outflow component and must be rotating, which
could lead to an additional change of the orbital structure, probably
of a similar order as derived here. 
The mass of the cloud directly estimated from observations depends on the
assumption of its structure. A change of the mass of the cloud 
can substantially affect hydrodynamical effects. Lowering
the cloud mass e.~g.~leads to a stronger ram pressure compression of the cloud and 
a shorter dissolution time scale due to hydrodynamical instabilities,
and hence influences the appearance of the cloud in the
position-velocity diagram. 
Furthermore, physical processes other than gravity and hydrodynamics
are neglected.  E.~g.~magnetic fields might change the dissolution time scale of the
G2 cloud as was recently simulated by \citet{McCourt_15a}.
This shows the importance of a careful comparison to
observations to derive possible parameters for the G2 cloud.

In the present paper, we largely omit the discussion of the post-pericenter
evolution of the cloud, as this necessitates a more detailed modelling 
of physical processes at work and
the detailed thermodynamic treatment of the gas
\citep[see e.~g.~][]{Burkert_12,Schartmann_12}.
The late-time
evolution also sensitively depends on the initial conditions of the
simulations \citep{Schartmann_12}. This is mainly caused by the steep profile of the
atmosphere at small radii where also the orbital evolution is fastest. 
Slight changes of the orbital evolution
of the cloud for various initial conditions due to the hydrodynamical
interaction can have strong effects on the post-pericenter evolution. 
Hence a more detailed understanding of the nature of the
cloud and the surrounding hot gas atmosphere 
is necessary for the prediction of the late-time evolution.   
Another challenge for the Compact Cloud scenario are the
L'-band observations \citep{Witzel_14} showing a constant intensity
and being consistent with a point source.
To take these into account in our Compact Cloud models would require a detailed
modelling of dust destruction processes in the hostile environment of
the Galactic Centre, which is beyond the scope of the current analysis.

Our simulations presented here are consistent
with the new interpretation of G2 being a condensation within a gas
streamer pointing towards Sgr~A* \citep{Pfuhl_15}. 
However, we still treat the cloud as being in direct contact with the
hot atmosphere. Whereas the pressure equilibrium assumption might be valid
(see discussion in \citealp{Burkert_12}), the development of fluid
instabilities at the boundary of the cloud will be
influenced when more or less comoving with a surrounding stream of
gas.

\section{Conclusions}
\label{sec:conclusions}

New simulations of the Compact Cloud Scenario are presented
to shed light on the origin and evolution of the dusty, ionized gas
cloud G2 on its way toward the massive BH in the Galactic
Center. We have updated our previous simulations \citep{Schartmann_12} in
two ways: we employ three-dimensional hydrodynamical
adaptive mesh refinement simulations to follow the evolution of the
cloud and we adapt the currently best-fit orbital solution based on
the cloud's Brackett-$\gamma$ emission \citep{Gillessen_13b}.

The primary aim of this study is to 
tackle the two main problems of the original Compact Cloud
  Scenario:
(1) the necessity of an in-situ starting point, which is unlikely, as
no source of mass could have been identified at the proposed position
and (2) the observed plateau in the Brackett-$\gamma$ light curves,
which has never been seen in simulations. 
To this end, we present a parameter study, in which we vary the starting
point of the cloud along G2's observed orbit, as well as 
resolution studies. The latter allow us to discuss 
possible interpretations of the observed Brackett-$\gamma$ evolution. 

From a detailed comparison of these hydrodynamical simulations with
the observed position-velocity diagrams as well as the Brackett-$\gamma$ 
light curves, we find that:
\begin{enumerate}
\item A starting point of the cloud within the range of the disks of
  young stars is favored by our models, which enables the interpretation of G2
  being the result of stellar processes. Possible candidate stars are
  S91 and IRS16 SW (see discussion in \citealp{Pfuhl_15} and \citet{Calderon_15}).
\item For starting times as early as this, we find that hydrodynamical
  effects slightly affect the orbital evolution of the cloud, which we
  correct for by applying a time shift. 
\item The problem is degenerate and depends sensitively on the main
  simulation parameters: cloud mass, profile of the atmosphere and
  further, mostly neglected, physical effects.  
  For the assumptions made in this publication, a reasonable
  adaptation of the observed PV diagrams can be reached for a starting
  time of approximately 1900.
\item We find that the resolution of the simulations critically affects the
    Brackett-$\gamma$ luminosity evolution, but not the spatially
    extended PV diagrams.
\item A detailed comparison with observations is
  vital to gain insight into the most important parameters  
  governing the evolution of the G2 cloud.
\end{enumerate}

One physically plausible interpretation in the
framework of these simulations is that the cloud is part of a clumpy
stream of gas pointing towards Sgr A* \citep{Pfuhl_15}. 
The {\it tail} component G2t, which we ignore in 
the analysis of
this publication, could
then be interpreted as a second condensation within this stream. 
In the case that G2 and the tail are unrelated \citep{Phifer_13,Valencia_15}, the
cloud could be interpreted as the result of a
collision of two stellar winds in the disk of young stars. 
This is consistent with our newly determined possible origin location of the
cloud(s). Depending on the characteristics of the involved stars, 
the shocked interstellar medium might reach densities high enough to 
trigger cooling instability, which then
leads to clump formation \citep{Burkert_12,Calderon_15}. 
Such a collision might as well efficiently redistribute angular
momentum, allowing a fraction of the clumps to end up on almost radial
trajectories, as is the case for G2.


\acknowledgments

We thank an anonymous referee for valuable comments.
This work was partly supported by the Deutsche Forschungsgemeinschaft priority program 1573
(``Physics of the Interstellar Medium").
Computer resources for this project have been provided by the 
Leibniz Supercomputing Center under grant h0075. 
The simulation analysis was partly performed on the swinSTAR
supercomputer at Swinburne University of Technology (Australia), utilising 
the yt-project \citep{Turk_11}. 


\bibliographystyle{apj}
\bibliography{apj-jour,literature}

\begin{thebibliography}{46}
\expandafter\ifx\csname natexlab\endcsname\relax\def\natexlab#1{#1}\fi

\bibitem[{{Anninos} {et~al.}(2012){Anninos}, {Fragile}, {Wilson}, \&
  {Murray}}]{Anninos_12}
{Anninos}, P., {Fragile}, P.~C., {Wilson}, J., \& {Murray}, S.~D. 2012, \apj,
  759, 132

\bibitem[{{Ballone} {et~al.}(2013){Ballone}, {Schartmann}, {Burkert},
  {Gillessen}, {Genzel}, {Fritz}, {Eisenhauer}, {Pfuhl}, \& {Ott}}]{Ballone_13}
{Ballone}, A., {Schartmann}, M., {Burkert}, A., {et~al.} 2013, \apj, 776, 13

\bibitem[{{Bartko} {et~al.}(2009){Bartko}, {Martins}, {Fritz}, {Genzel},
  {Levin}, {Perets}, {Paumard}, {Nayakshin}, {Gerhard}, {Alexander},
  {Dodds-Eden}, {Eisenhauer}, {Gillessen}, {Mascetti}, {Ott}, {Perrin},
  {Pfuhl}, {Reid}, {Rouan}, {Sternberg}, \& {Trippe}}]{Bartko_09}
{Bartko}, H., {Martins}, F., {Fritz}, T.~K., {et~al.} 2009, \apj, 697, 1741

\bibitem[{{Bonnet} {et~al.}(2004){Bonnet}, {Abuter}, {Baker}, {Bornemann},
  {Brown}, {Castillo}, {Conzelmann}, {Damster}, {Davies}, {Delabre},
  {Donaldson}, {Dumas}, {Eisenhauer}, {Elswijk}, {Fedrigo}, {Finger},
  {Gemperlein}, {Genzel}, {Gilbert}, {Gillet}, {Goldbrunner}, {Horrobin}, {Ter
  Horst}, {Huber}, {Hubin}, {Iserlohe}, {Kaufer}, {Kissler-Patig}, {Kragt},
  {Kroes}, {Lehnert}, {Lieb}, {Liske}, {Lizon}, {Lutz}, {Modigliani}, {Monnet},
  {Nesvadba}, {Patig}, {Pragt}, {Reunanen}, {R{\"o}hrle}, {Rossi}, {Schmutzer},
  {Schoenmaker}, {Schreiber}, {Stroebele}, {Szeifert}, {Tacconi}, {Tecza},
  {Thatte}, {Tordo}, {van der Werf}, \& {Weisz}}]{Bonnet_04}
{Bonnet}, H., {Abuter}, R., {Baker}, A., {et~al.} 2004, The Messenger, 117, 17

\bibitem[{{Bower} {et~al.}(2015){Bower}, {Markoff}, {Dexter}, {Gurwell},
  {Moran}, {Brunthaler}, {Falcke}, {Fragile}, {Maitra}, {Marrone}, {Peck},
  {Rushton}, \& {Wright}}]{Bower_15}
{Bower}, G.~C., {Markoff}, S., {Dexter}, J., {et~al.} 2015, \apj, 802, 69

\bibitem[{{Burkert} {et~al.}(2012){Burkert}, {Schartmann}, {Alig}, {Gillessen},
  {Genzel}, {Fritz}, \& {Eisenhauer}}]{Burkert_12}
{Burkert}, A., {Schartmann}, M., {Alig}, C., {et~al.} 2012, \apj, 750, 58

\bibitem[{{Calder{\'o}n} {et~al.}(2015){Calder{\'o}n}, {Ballone}, {Cuadra},
  {Schartmann}, {Burkert}, \& {Gillessen}}]{Calderon_15}
{Calder{\'o}n}, D., {Ballone}, A., {Cuadra}, J., {et~al.} 2015, ArXiv e-prints

\bibitem[{{Cuadra} {et~al.}(2006){Cuadra}, {Nayakshin}, {Springel}, \& {Di
  Matteo}}]{Cuadra_06}
{Cuadra}, J., {Nayakshin}, S., {Springel}, V., \& {Di Matteo}, T. 2006, \mnras,
  366, 358

\bibitem[{{De Colle} {et~al.}(2014){De Colle}, {Raga}, {Contreras-Torres}, \&
  {Toledo-Roy}}]{Colle_14}
{De Colle}, F., {Raga}, A.~C., {Contreras-Torres}, F.~F., \& {Toledo-Roy},
  J.~C. 2014, \apjl, 789, L33

\bibitem[{{Eatough} {et~al.}(2013){Eatough}, {Falcke}, {Karuppusamy}, {Lee},
  {Champion}, {Keane}, {Desvignes}, {Schnitzeler}, {Spitler}, {Kramer},
  {Klein}, {Bassa}, {Bower}, {Brunthaler}, {Cognard}, {Deller}, {Demorest},
  {Freire}, {Kraus}, {Lyne}, {Noutsos}, {Stappers}, \& {Wex}}]{Eatough_13}
{Eatough}, R.~P., {Falcke}, H., {Karuppusamy}, R., {et~al.} 2013, \nat, 501,
  391

\bibitem[{{Eisenhauer} {et~al.}(2003){Eisenhauer}, {Abuter}, {Bickert},
  {Biancat-Marchet}, {Bonnet}, {Brynnel}, {Conzelmann}, {Delabre}, {Donaldson},
  {Farinato}, {Fedrigo}, {Genzel}, {Hubin}, {Iserlohe}, {Kasper},
  {Kissler-Patig}, {Monnet}, {Roehrle}, {Schreiber}, {Stroebele}, {Tecza},
  {Thatte}, \& {Weisz}}]{Eisenhauer_03}
{Eisenhauer}, F., {Abuter}, R., {Bickert}, K., {et~al.} 2003, in Society of
  Photo-Optical Instrumentation Engineers (SPIE) Conference Series, Vol. 4841,
  Instrument Design and Performance for Optical/Infrared Ground-based
  Telescopes, ed. M.~{Iye} \& A.~F.~M. {Moorwood}, 1548--1561

\bibitem[{{Ferland}(1980)}]{Ferland_80}
{Ferland}, G.~J. 1980, \pasp, 92, 596

\bibitem[{{Ghez} {et~al.}(2005){Ghez}, {Hornstein}, {Lu}, {Bouchez}, {Le
  Mignant}, {van Dam}, {Wizinowich}, {Matthews}, {Morris}, {Becklin},
  {Campbell}, {Chin}, {Hartman}, {Johansson}, {Lafon}, {Stomski}, \&
  {Summers}}]{Ghez_05}
{Ghez}, A.~M., {Hornstein}, S.~D., {Lu}, J.~R., {et~al.} 2005, \apj, 635, 1087

\bibitem[{{Ghez} {et~al.}(2008){Ghez}, {Salim}, {Weinberg}, {Lu}, {Do}, {Dunn},
  {Matthews}, {Morris}, {Yelda}, {Becklin}, {Kremenek}, {Milosavljevic}, \&
  {Naiman}}]{Ghez_08}
{Ghez}, A.~M., {Salim}, S., {Weinberg}, N.~N., {et~al.} 2008, \apj, 689, 1044

\bibitem[{{Gillessen} {et~al.}(2009){Gillessen}, {Eisenhauer}, {Trippe},
  {Alexander}, {Genzel}, {Martins}, \& {Ott}}]{Gillessen_09}
{Gillessen}, S., {Eisenhauer}, F., {Trippe}, S., {et~al.} 2009, \apj, 692, 1075

\bibitem[{{Gillessen} {et~al.}(2012){Gillessen}, {Genzel}, {Fritz}, {Quataert},
  {Alig}, {Burkert}, {Cuadra}, {Eisenhauer}, {Pfuhl}, {Dodds-Eden}, {Gammie},
  \& {Ott}}]{Gillessen_12}
{Gillessen}, S., {Genzel}, R., {Fritz}, T.~K., {et~al.} 2012, \nat, 481, 51

\bibitem[{{Gillessen} {et~al.}(2013{\natexlab{a}}){Gillessen}, {Genzel},
  {Fritz}, {Eisenhauer}, {Pfuhl}, {Ott}, {Cuadra}, {Schartmann}, \&
  {Burkert}}]{Gillessen_13a}
---. 2013{\natexlab{a}}, \apj, 763, 78

\bibitem[{{Gillessen} {et~al.}(2013{\natexlab{b}}){Gillessen}, {Genzel},
  {Fritz}, {Eisenhauer}, {Pfuhl}, {Ott}, {Schartmann}, {Ballone}, \&
  {Burkert}}]{Gillessen_13b}
---. 2013{\natexlab{b}}, \apj, 774, 44

\bibitem[{{Guillochon} {et~al.}(2014){Guillochon}, {Loeb}, {MacLeod}, \&
  {Ramirez-Ruiz}}]{Guillochon_14}
{Guillochon}, J., {Loeb}, A., {MacLeod}, M., \& {Ramirez-Ruiz}, E. 2014, \apjl,
  786, L12

\bibitem[{{Hamann} \& {Ferland}(1999)}]{Hamann_99}
{Hamann}, F., \& {Ferland}, G. 1999, \araa, 37, 487

\bibitem[{{Larkin} {et~al.}(2006){Larkin}, {Barczys}, {Krabbe}, {Adkins},
  {Aliado}, {Amico}, {Brims}, {Campbell}, {Canfield}, {Gasaway}, {Honey},
  {Iserlohe}, {Johnson}, {Kress}, {LaFreniere}, {Magnone}, {Magnone},
  {McElwain}, {Moon}, {Quirrenbach}, {Skulason}, {Song}, {Spencer}, {Weiss}, \&
  {Wright}}]{Larkin_06}
{Larkin}, J., {Barczys}, M., {Krabbe}, A., {et~al.} 2006, \nar, 50, 362

\bibitem[{{Lu} {et~al.}(2009){Lu}, {Ghez}, {Hornstein}, {Morris}, {Becklin}, \&
  {Matthews}}]{Lu_09}
{Lu}, J.~R., {Ghez}, A.~M., {Hornstein}, S.~D., {et~al.} 2009, \apj, 690, 1463

\bibitem[{{McCourt} \& {Madigan}(2015)}]{McCourt_15b}
{McCourt}, M., \& {Madigan}, A.-M. 2015, ArXiv e-prints

\bibitem[{{McCourt} {et~al.}(2015){McCourt}, {O'Leary}, {Madigan}, \&
  {Quataert}}]{McCourt_15a}
{McCourt}, M., {O'Leary}, R.~M., {Madigan}, A.-M., \& {Quataert}, E. 2015,
  \mnras, 449, 2

\bibitem[{{Meyer} \& {Meyer-Hofmeister}(2012)}]{Meyer_12}
{Meyer}, F., \& {Meyer-Hofmeister}, E. 2012, \aap, 546, L2

\bibitem[{{Mignone} {et~al.}(2007){Mignone}, {Bodo}, {Massaglia}, {Matsakos},
  {Tesileanu}, {Zanni}, \& {Ferrari}}]{Mignone_07}
{Mignone}, A., {Bodo}, G., {Massaglia}, S., {et~al.} 2007, \apjs, 170, 228

\bibitem[{{Mignone} {et~al.}(2012){Mignone}, {Zanni}, {Tzeferacos}, {van
  Straalen}, {Colella}, \& {Bodo}}]{Mignone_12}
{Mignone}, A., {Zanni}, C., {Tzeferacos}, P., {et~al.} 2012, \apjs, 198, 7

\bibitem[{{Miralda-Escud{\'e}}(2012)}]{Miralda_Escude_12}
{Miralda-Escud{\'e}}, J. 2012, \apj, 756, 86

\bibitem[{{Murray-Clay} \& {Loeb}(2012)}]{Murray_Clay_12}
{Murray-Clay}, R.~A., \& {Loeb}, A. 2012, Nature Communications, 3

\bibitem[{{Osterbrock} \& {Ferland}(2006)}]{Osterbrock_06}
{Osterbrock}, D.~E., \& {Ferland}, G.~J. 2006, {Astrophysics of gaseous nebulae
  and active galactic nuclei} (University Science Books)

\bibitem[{{Park} {et~al.}(2015){Park}, {Trippe}, {Krichbaum}, {Kim}, {Kino},
  {Bertarini}, {Bremer}, \& {de Vicente}}]{Park_15}
{Park}, J.-H., {Trippe}, S., {Krichbaum}, T.~P., {et~al.} 2015, \aap, 576, L16

\bibitem[{{Paumard} {et~al.}(2006){Paumard}, {Genzel}, {Martins}, {Nayakshin},
  {Beloborodov}, {Levin}, {Trippe}, {Eisenhauer}, {Ott}, {Gillessen}, {Abuter},
  {Cuadra}, {Alexander}, \& {Sternberg}}]{Paumard_06}
{Paumard}, T., {Genzel}, R., {Martins}, F., {et~al.} 2006, \apj, 643, 1011

\bibitem[{{Pfuhl} {et~al.}(2015){Pfuhl}, {Gillessen}, {Eisenhauer}, {Genzel},
  {Plewa}, {Ott}, {Ballone}, {Schartmann}, {Burkert}, {Fritz}, {Sari},
  {Steinberg}, \& {Madigan}}]{Pfuhl_15}
{Pfuhl}, O., {Gillessen}, S., {Eisenhauer}, F., {et~al.} 2015, \apj, 798, 111

\bibitem[{{Phifer} {et~al.}(2013){Phifer}, {Do}, {Meyer}, {Ghez}, {Witzel},
  {Yelda}, {Boehle}, {Lu}, {Morris}, {Becklin}, \& {Matthews}}]{Phifer_13}
{Phifer}, K., {Do}, T., {Meyer}, L., {et~al.} 2013, \apjl, 773, L13

\bibitem[{{Ponti} {et~al.}(2015){Ponti}, {De Marco}, {Morris}, {Merloni},
  {Munoz-Darias}, {Clavel}, {Haggard}, {Zhang}, {Nandra}, {Gillessen}, {Mori},
  {Neilsen}, {Rea}, {Degenaar}, {Terrier}, \& {Goldwurm}}]{Ponti_15}
{Ponti}, G., {De Marco}, B., {Morris}, M.~R., {et~al.} 2015, ArXiv e-prints

\bibitem[{{Prodan} {et~al.}(2015){Prodan}, {Antonini}, \& {Perets}}]{Prodan_15}
{Prodan}, S., {Antonini}, F., \& {Perets}, H.~B. 2015, \apj, 799, 118

\bibitem[{{Schartmann} {et~al.}(2012){Schartmann}, {Burkert}, {Alig},
  {Gillessen}, {Genzel}, {Eisenhauer}, \& {Fritz}}]{Schartmann_12}
{Schartmann}, M., {Burkert}, A., {Alig}, C., {et~al.} 2012, \apj, 755, 155

\bibitem[{{Scoville} \& {Burkert}(2013)}]{Scoville_13}
{Scoville}, N., \& {Burkert}, A. 2013, \apj, 768, 108

\bibitem[{{Shcherbakov}(2014)}]{Shcherbakov_14}
{Shcherbakov}, R.~V. 2014, \apj, 783, 31

\bibitem[{{Turk} {et~al.}(2011){Turk}, {Smith}, {Oishi}, {Skory}, {Skillman},
  {Abel}, \& {Norman}}]{Turk_11}
{Turk}, M.~J., {Smith}, B.~D., {Oishi}, J.~S., {et~al.} 2011, \apjs, 192, 9

\bibitem[{{Valencia-S.} {et~al.}(2015){Valencia-S.}, {Eckart}, {Zaja{\v c}ek},
  {Peissker}, {Parsa}, {Grosso}, {Mossoux}, {Porquet}, {Jalali}, {Karas},
  {Yazici}, {Shahzamanian}, {Sabha}, {Saalfeld}, {Smajic}, {Grellmann},
  {Moser}, {Horrobin}, {Borkar}, {Garc{\'{\i}}a-Mar{\'{\i}}n}, {Dov{\v c}iak},
  {Kunneriath}, {Karssen}, {Bursa}, {Straubmeier}, \& {Bushouse}}]{Valencia_15}
{Valencia-S.}, M., {Eckart}, A., {Zaja{\v c}ek}, M., {et~al.} 2015, \apj, 800,
  125

\bibitem[{{van Dam} {et~al.}(2006){van Dam}, {Bouchez}, {Le Mignant},
  {Johansson}, {Wizinowich}, {Campbell}, {Chin}, {Hartman}, {Lafon}, {Stomski},
  \& {Summers}}]{vanDam_06}
{van Dam}, M.~A., {Bouchez}, A.~H., {Le Mignant}, D., {et~al.} 2006, \pasp,
  118, 310

\bibitem[{{Witzel} {et~al.}(2014){Witzel}, {Ghez}, {Morris}, {Sitarski},
  {Boehle}, {Naoz}, {Campbell}, {Becklin}, {Canalizo}, {Chappell}, {Do}, {Lu},
  {Matthews}, {Meyer}, {Stockton}, {Wizinowich}, \& {Yelda}}]{Witzel_14}
{Witzel}, G., {Ghez}, A.~M., {Morris}, M.~R., {et~al.} 2014, \apjl, 796, L8

\bibitem[{{Wizinowich} {et~al.}(2006){Wizinowich}, {Le Mignant}, {Bouchez},
  {Campbell}, {Chin}, {Contos}, {van Dam}, {Hartman}, {Johansson}, {Lafon},
  {Lewis}, {Stomski}, {Summers}, {Brown}, {Danforth}, {Max}, \&
  {Pennington}}]{Wizinowich_06}
{Wizinowich}, P.~L., {Le Mignant}, D., {Bouchez}, A.~H., {et~al.} 2006, \pasp,
  118, 297

\bibitem[{{Yelda} {et~al.}(2014){Yelda}, {Ghez}, {Lu}, {Do}, {Meyer}, {Morris},
  \& {Matthews}}]{Yelda_14}
{Yelda}, S., {Ghez}, A.~M., {Lu}, J.~R., {et~al.} 2014, \apj, 783, 131

\bibitem[{{Yuan} {et~al.}(2003){Yuan}, {Quataert}, \& {Narayan}}]{Yuan_03}
{Yuan}, F., {Quataert}, E., \& {Narayan}, R. 2003, \apj, 598, 301

\end{thebibliography}


\end{document}